\documentclass[twocolumns]{aa}

\usepackage{graphicx}
\def\gs{{_>\atop^{\sim}}}

\begin{document}
\title{The faint X-ray Source
Population near 3C 295}

\author{Valerio D'Elia$^1$, Fabrizio Fiore$^1$, Martin Elvis$^2$,
  Massimo Cappi$^3$, Smita Mathur$^4$, Pasquale Mazzotta$^5$, Emilio Falco$^6$, Filomena Cocchia$^1$
}
\institute {INAF-Osservatorio Astronomico di Roma \\
via Frascati 33, Monteporzio-Catone (RM), I00040 Italy.
\and
Harvard-Smithsonian Center for Astrophysics, 
Cambridge MA 02138, USA
\and
IASF-CNR, Bologna, Italy
\and
The Ohio State University, Columbus, OH 43210, USA.
\and
University of Durham South Road Durham DH1 3LE, UK
\and
Smithsonian Institution Whipple Observatory  AZ 85645 USA
}
\date{July 30 2003}

\abstract { We present a statistical analysis of the {\it Chandra}
observation of the source field around the 3C 295 galaxy cluster
(z=0.46) aimed at the search for clustering of X-ray sources. 
We applied three different methods of analysis, all suggesting a
strong clustering in the field on scales of a few arcmin. In
particular 1) the logN-logS computed separately for the four ACIS-I
chips reveals that there is a significant ($3.2\; \sigma$ in the
$0.5-2$ keV, $3.3\; \sigma$ in the $2-10$ keV and $4.0\; \sigma$ in
the $0.5-10$ keV band) excess of sources to the North-North East and a
void to the South of the central cluster.  2) the two point,
two-dimensional Kolmogorov-Smirnov (KS) test, shows the probability
that the sources are uniformly distributed is only a few percent. 3) a
strong spatial correlation emerges from the study of the angular
correlation function of the field: the angular correlation function
(ACF) shows a clear signal on scales of $0.5\div 5$ arcmin,
correlation angle in the $0.5-7$ keV band $\theta_0=8.5^{+6.5}_{-4.5}$, $90$\% 
confidence limit
(assuming a power law ACF with
slope $\gamma=1.8$). This correlation angle is $2$ times higher than that 
of a sample of $8$ ACIS-I field at the $2.5 \; \sigma$ confidence level.
The above scales translate to 0.2$\div$2
Mpc at the cluster redshift, higher than the typical cluster core
radius, and more similar to the size of a ``filament'' of the large
scale structure.


\authorrunning {D'Elia et al.}
\titlerunning {X-ray sources in the 3C295 field}
}
\maketitle

\section{Introduction}

N-body and hydrodynamical simulations show that clusters of
galaxies lie at the nexus of several filaments of galaxies (see
e.g. Peacock 1999, Dav\`e et al. 2001 and references therein).  Such
filaments map out the ``cosmic web'' of voids and filaments of the
large scale structure (LSS) of the Universe. According to the same
simulations these filaments contain a large fraction ($30-40\%$) of
the baryons in the Universe at z$<1$ (the remainder ending up in the
hot gas in clusters of galaxies on one side, and in stars and cold gas
clouds on the other side).  Despite its larger total mass,
observations of the intergalactic matter in filaments have yielded so
far only limited information, mostly due to its low density (most of
the baryons in this phase should be at densities only 10-100 times
higher than the average density in the Universe).  The most direct
ways to detect a filament at low redshift 
is through its soft X-ray diffuse emission
(see e.g. Zappacosta et al. 2002, Soltan, Freyberg \& Hasinger 2002), 
or through soft X-ray and UV
absorption line studies (see e.g. Fiore et al. 2000 and references
therein, Nicastro et al. 2002, Nicastro et al. 2003, Mathur et
al. 2003).  Both methods require very difficult observations, at the
limit of the present generation of X-ray and UV facilities.
Alternatively, filaments could be mapped out by galaxies (Daddi et al. 2001, 
Giavalisco \& Dickinson 2001) and by the
much more luminous Active Galactic Nuclei (AGNs), assuming that AGNs
trace galaxies. Since rich clusters of galaxies are good
indicators of regions of sky where filaments converge, numerous AGN
searches around clusters of galaxies have been performed in the past.
Several of these studies suggest overdensities of AGNs around distant
clusters of galaxies (Molnar et al. 2002 for the cluster Abell 1995;
Best et al. 2002 for MS1054-03; Martini et al. 2002 for Abell 2104;
Pentericci et al. 2002 for the protocluster at $z \sim 2.16$ around
the radio galaxy MRC 1138-206; see also Almaini et al. 2003 for the
ELAIS North field). Many of these studies have been performed in X-rays,
since extragalactic X-ray sources, which are mostly AGNs, have a space density $\sim
10$ times higher than optically selected AGNs (see Yang et al. 2003), 
and therefore provide denser tracing of LSS.  One of the first studies 
of the X-ray source population in cluster fields was performed by Cappi et al.
(2001) who studied the {\it Chandra} $8\times 8$ arcmin ACIS fields around
the distant cluster of galaxies RX J003033.2+261819 ($z=0.5$) and 3C
295 ($z=0.46$).  Cappi et al. (2001) reported the tentative detection of an
overdensity of faint X-ray sources in a region of a few arcmin around
both clusters, with respect to the average X-ray source density at the
same flux limit.  However, the observations were too short ($\sim 30$
ks and $\sim 18$ ks, respectively), and the source samples
consequently too small, to derive any more detailed conclusion.

In addition to counting sources in selected sky areas, the clustering
of X-ray sources can be studied using other observational and
statistical tools.  Gilli et al. (2003) report narrow spikes in the
redshift distribution of the sources in the {\it Chandra} Deep Field South
(CDFS), indicating strong clustering of sources in these narrow
redshift ranges.  If the source samples are sufficiently large (of the
order of 100 sources or more) one derives more detailed and
quantitative information on the source clustering by studying the
angular correlation function (ACF) of the sources in the
field. Vikhlinin \& Forman (1995) were among the first to study
the ACF in the X-ray band; they evaluated an average ACF from a large
set of deep ROSAT observations, covering in total 40 deg$^2$ of sky.
They found positive correlation on scales from a fraction of arcmin to
tens of arcmin.  However, their best fit ``correlation angle''
$\theta_0 \sim 10$ arcsec, is smaller than the ROSAT PSPC Point Spread
Function ($\sim 25$ arcsec FWHM on-axis); as noted by the authors,
this leads to an ``amplification bias'' and their measured ACF is
consequently somewhat overestimated.  

This effect can be greatly mitigated by
using a telescope like {\it Chandra}, whose on-axis PSF
is only 0.5 arcsec FWHM, 50 times better than the PSPC on-axis PSF.
So far a correlation analysis of {\it Chandra} X-ray sources have been
published only by Giacconi et al. (2000), who used the first 100 ks of
observation of the CDFS, and by Yang et al. (2003) who analyzed a
mosaic of 9 moderately deep (30ks) {\it Chandra} pointings of the
Lockman Hole area.  Yang et al. could study the source angular
correlation on scales of several tens of arcmin, being limited on
smaller scales by the small number of detected sources per unit area.

Our main goal in this paper is to push this kind of analysis toward a)
smaller scales, of order of a few arcmin; and b) targeting a field
where the likelihood of observing LSS is high, i.e. a field
around a cluster of galaxies.  For these reasons we observed again
with {\it Chandra} for about 100ks the 3C295 field, where Cappi et
al. (2001) found tentative evidence of an overdensity of faint X-ray
sources in a region of a few arcmin around the central cluster.

The paper is organized as follows: Section 2 presents the observations
and data reduction; Section 3 presents the results of our analysis;
Section 4 discusses such results and draws our conclusions.

\section{{\it Chandra} observations and data reduction}

{\it Chandra} observed the $16'\times 16'$ field around the 3C 295
cluster with ACIS-I (Garmire 1997) on 2001, May 18. The aim point
of the observation is located at the position of 3C 295 $\alpha =$
14:11:10, $\delta=+$52:13:01 (J2000) and exposure time is $\sim 92$
ks. The data reduction was carried out using the {\it Chandra}
Interactive Analysis of Observations (CIAO) software version 2.1.3
(http://cxc.harvard.edu/ciao). No strong background flares are 
present in the observation. All the data analysis has been
performed separately in the soft $0.5-2$ keV band, in the hard $2-7$
keV band and in the broad $0.5-7$ keV band. Since the ACIS background
rises sharply above $\sim 7$ keV, we excluded all data with $E > 7$
keV.

In order to check our analysis method, an 
identical analysis was performed for the {\it
Chandra} Deep Field South (CDFS, see Giacconi et al. 2002) in the
$0.5-2$ keV and $2-7$ keV bands. The
CDFS is a collection of 11 separate ACIS-I observations centered on
the same point, $\alpha =$03:32:28.0, $\delta=-$27:48:30 (J2000), but
with different roll angles. For this reason, the effective exposure
time changes in different regions of the field. For the CDFS analysis
we selected only the region with the maximal effective exposure time,
which is $\sim 942$ ks; a region that is slightly smaller than the
full 3C 295 field.

In the following subsection the source detection procedure is
described, while subsection 2.2. deals with the sky coverage
evaluation, which will be used in section 3 for the logN-logS
calculation.

\subsection{Source detections}

The source detection was carried out using the `PWDetect'
algorithm. This is a wavelet-based source detection code for {\it
Chandra} X-ray data developed at the Osservatorio Astronomico di
Palermo (Damiani et al. 1997a, 1997b).  PWDetect performs a multiscale
analysis of the data, allowing detection of both pointlike and
moderately extended sources in the entire field of view.  This
detection algorithm was compared with other detection tools, namely,
the CIAO detection tool {\sc celldetect} and the Ximage {\sc
detect}. The latter algorithms detected a smaller number of sources
than PWDetect in both the CDFS and 3C 295 field, and in all energy
bands. In addition, they tended to find extended sources as multiple
detections.

\begin{figure}
\centering
\includegraphics[angle=0,width=8cm]{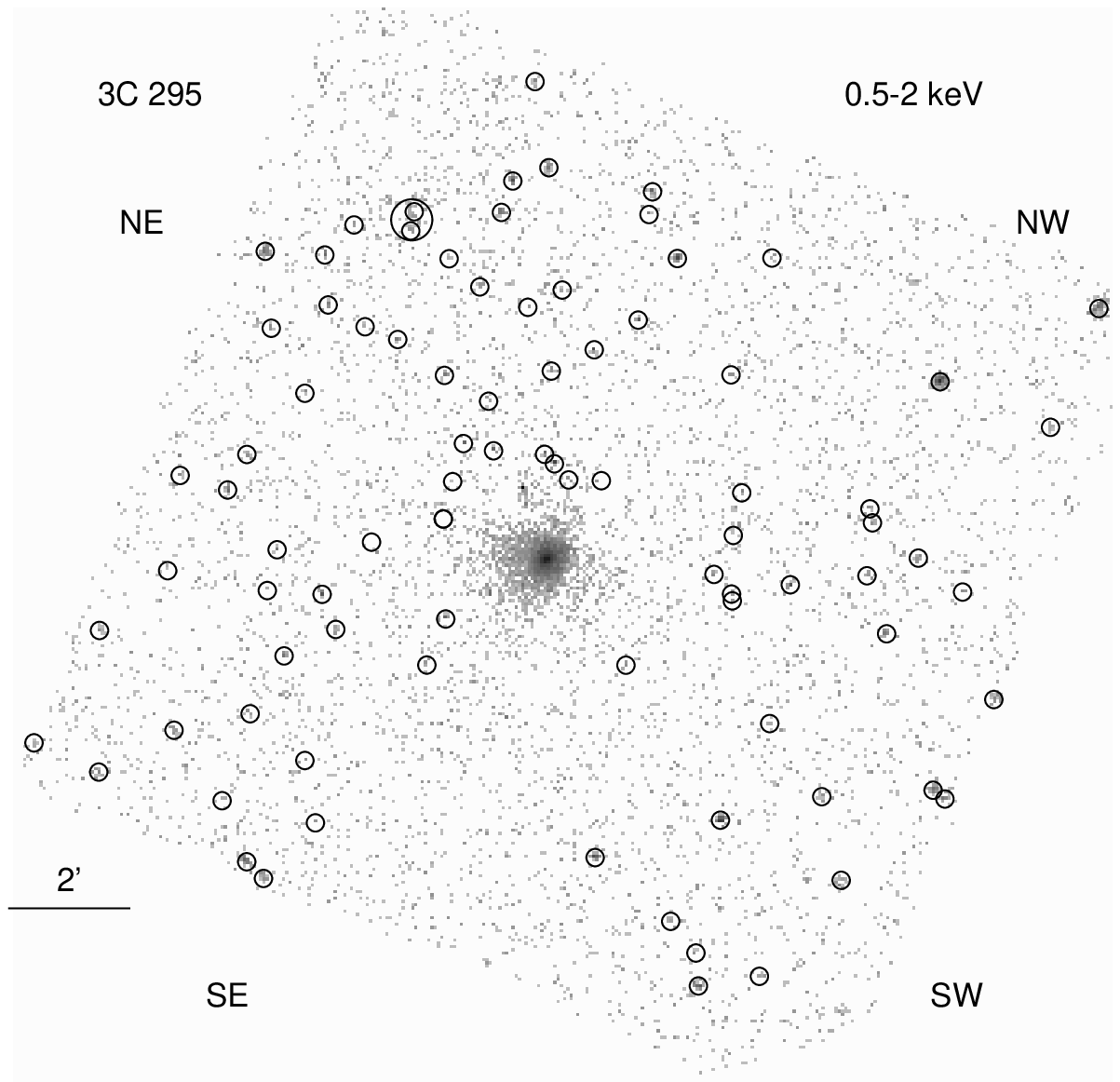}
\includegraphics[angle=0,width=8.5cm]{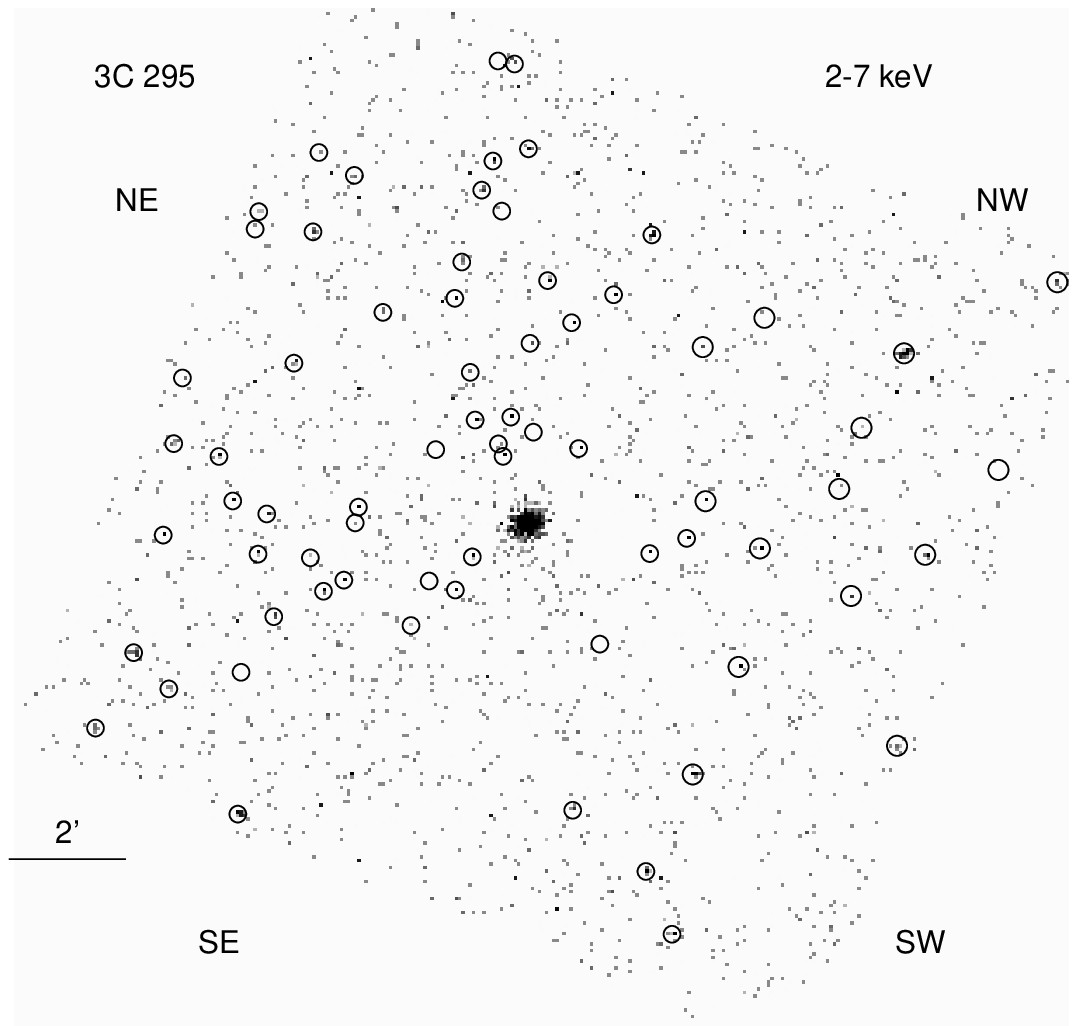}
\includegraphics[angle=0,width=7.5cm]{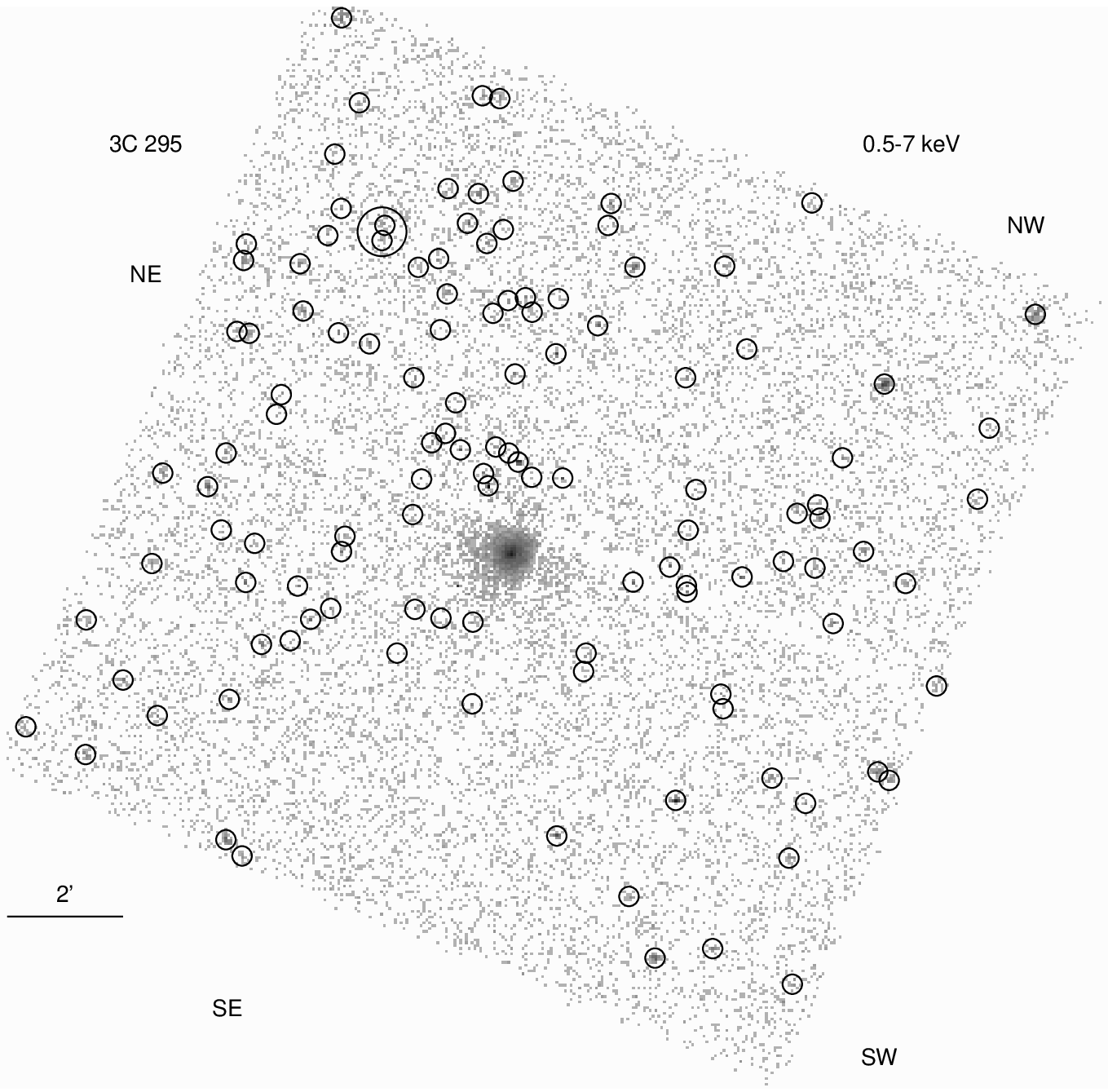}
\caption{ The {\it Chandra} 3C 295 field in the $0.5-2$ keV band (upper
panel), in the $2-7$ keV band (central panel) and in the $0.5-7$ keV
band (bottom panel).  Circles represent the sources detected with the
wavelet-based source detection code `PWdetect'. The largest circles in
the upper and bottom panels indicates a very extended source, possibly
a group of galaxies.  The brightest source in the center of the field
is the cluster of galaxies 3C 295.  }
\label{spe1}
\end{figure}

PWDetect requires a {\it Chandra} FITS event file and the associated
exposure map as inputs. The exposure maps were computed using the CIAO
task {\sc mkexpmap} for each ACIS-I chip and then combined together
into the single exposure map required by PWDetect for each of the
$0.5-2$ keV, $2-7$ keV and $0.5-7$ keV energy bands.  The sources were
detected using minimum and maximum wavelet scales of 0.5 and 16
arcsec, respectively. 
We assumed a probability detection threshold of $2\times10^{-5}$. 
Since the local background level is of the order of $0.1$ 
cts/arcsec$^2$, the minimum number of counts for a source to
be detected is $\sim 7$. The assumed probability and the number
of independent cells used by PWdetect give the expected number of
spurios sources per field; this number results to be $< 1$ in all
energy bands.
%
%
%
We identified 89 sources in the $0.5-2$ keV band lying
within the flux range $3.2\times 10^{-16} - 4.4\times 10^{-13}$ ergs
cm$^{-2}$ s$^{-1}$, 71 sources in the $2-7$ keV band within the flux
range $1.7\times 10^{-15} - 5.8\times 10^{-14}$ ergs cm$^{-2}$
s$^{-1}$ and $121$ sources in the $0.5-7$ keV band in the flux range
$8.8\times 10^{-16} - 1.2\times 10^{-13}$ ergs cm$^{-2}$ s$^{-1}$.
The central cluster was not considered in the analysis, thus we
excluded the detections in the $\sim 2' \times 2'$ central region.
The counts in the $0.5-2$ keV, $2-7$ keV and $0.5-7$ keV bands were
converted in $0.5-2$ keV, $2-10$ keV and $0.5-10$ keV fluxes,
respectively, by multipling the counts by the conversion factors
$5.65\times 10^{-12}$, $2.70 \times10^{-11}$ and $1.24 \times10^{-11}$
(from PIMMS: http://heasarc.gsfc.nasa.gov/Tools/w3pimms.html).  These
factors are appropriate for a $\Gamma=1.8$ power law spectrum with a
Galactic absorption toward the 3C 295 field of $N_H = 1.33 \times
10^{20}$ cm$^{-2}$ (Dickey \& Lockman 1990), and take into account the
quantum efficiency degradation of the ACIS chips
(http://asc.harvard.edu/cal/Acis).

Fig. 1 shows the 3C 295 field in the first two bands; the detected
sources are indicated by circles. The larger circle in the upper left
chip of the $0.5-2$ keV picture indicates an extended source (detected
on a $16$ arcsec scale, see next subsection) which is  
probably a group 
or a cluster of galaxies. A close observation of this region shows other
count peaks which could be other X-ray sources, possibly the galaxies in 
the group, below the detection threshold, althouh they may be spurious
detections caused by positive fluctuations of the surface brightness.

Visually the sources give a strong impression of being concentrated to
the East, and expecially the NE. This is the same trend reported in
Cappi et al. (2001). We will proceed to quantify this impression in the
following sections.

\subsection{Extended sources}

As mentioned before, in the soft and broad band, we identified an
extended source on a scale of $16$ arcsec, which could also be 
two or more confused point sources. This source is marked 
in fig. 1 by a large circle and its position is ra $= 14:11:37$, dec $= +52:18:35$; its  
$0.5-2$ keV flux is $ 4.4\times 10^{-15}$ ergs cm$^{-2}$ s$^{-1}$ 
and the $0.5-7$ keV one is $ = 1.1\times 10^{-14}$ ergs cm$^{-2}$ s$^{-1}$.
In fig. 2 we show the contours of the $0.5- 7 $ keV image superimposed
to the image taken at the TNG with the DOLORES camera on march 19,
2002, with an exposure time of $900$ s; the seeing was $1.3$ arcsec
and the limiting magnitude $R_{lim}= 24.1$. 
The southern strong X-ray emitter, which is a pointlike source, 
has no optical counterpart down to the limiting magnitude  
of the observation, while the extended emission is centered on two
bright $R_{mag} \sim 18.5$ galaxies.
\begin{figure}
\centering
\includegraphics[angle=0,width=8cm]{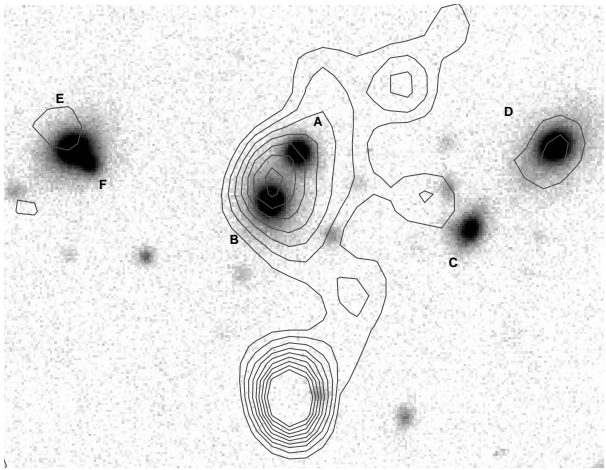}
\caption{The contours of the $0.5-7$ keV band image overimposed to the 
optical (R band) image of the extended source identified 
in the soft and broad band. Capital letters are associated
to galaxies whose R observed magnitudes are reported in tab. 1.The image
width is $\sim 1.25$ arcmin}
\label{spe1}
\end{figure}
Tab. 1 reports the magnitudes of the galaxies marked with capital letters 
in Fig. 2.

\begin{table*}[ht]
\caption{\bf Observed optical magnitudes in the R band of the sources in fig. 2}
\footnotesize
\smallskip
\begin{tabular}{|l|c|}
\hline
sources & $R_{mag}$\\ 
\hline
A&$18.8$\\
\hline
B&$18.15$\\
\hline
C&$19.11$\\
\hline
D&$17.95$\\
\hline
E&$18.18$\\
\hline
F&$19.11$\\
\hline
\end{tabular}

\normalsize
\end{table*}

If we assume that these galaxies are at a $z$ similar to that
of 3C 295, $z =0.46$, then their absolute magnitude is $M_B \simeq -23$.
This is almost two orders of magnitude more luminous than the break
magnitude of the galaxy luminosity function at $z = 0.5$, that is 
$M_B^* = - 21.3$ (e.g. Poli et al. 2003). 
Moreover, we note that the central radio galaxy of the 3C 295 cluster
has $R_{mag} = 18.4$ in our image, while the X-ray flux of the 3C 295 cluster is
$\sim 1.0 \times 10^{-12}$ ergs cm$^{-2}$ s$^{-1}$ in the $0.5 - 4.5$ keV (see e.g. 
Henry \& Henriksen ). We conclude that it is unlikely 
that the  optical sources corresponding to the extended x-ray emission are located 
at $z = 0.46$. 
Since the extended source is not detected in the $2-7$ keV band, we can only put
an upper limit on the temperature of the putative cluster or group; using 
a Raymond-Smith model with solar abundances ratio of  $0.4$, we find 
$T < 3$ keV if we assume the same redshift of 3C 295 and $T < 2$ keV for 
$z = 0$. On the other hand, an upper temperature of $2\div 3 $ keV poses 
an upper limit on the luminosity and thus on the redshift; unfortunately,
this limit is not significant for our purposes.  
 
We checked for other signs of diffuse emission in the 3C 295 field, and
in particular in the NE region, where the detected sources appear to be concentrated.
To do this, we subtracted the detected sources from the image and we 
computed the average background level of the observation in the  $0.5-7$ keV band; 
the central
part of the image was excluded because of the high level of counts due to
the central cluster. Next, we compared this background level with the 
counts evaluated in 25 regions of the field. The area of the regions 
has been chosen to vary in the range $10^3 \div 10^4$ arcsec$^2$. The excess with 
respect to the background, if any, is never significant at $> 3\;\sigma$
confidence level. Counts were converted to fluxes assuming a Raymond-Smith model
with $T = 1$ keV, and a solar abundances ratio of $0.4$. 
The $3\;\sigma$ upper limit on the flux  
is then $2.7 \times 10^{-14}$ ergs cm$^{-2}$ s$^{-1}$ arcmin$^{-2}$. Thus we can 
conclude that any diffuse emission is too faint to be detected.

\subsection{Sky coverage}

`Sky coverage' defines the area of the sky sensitive to a given 
flux limit,
as a function of flux. Due to vignetting effects and to the increase
of the size of the PSF in the outer regions of the detector, the
center of the ACIS-I field of view is the most sensitive: i.e. for a
given exposure time, the central part of the detector is able to
detect fainter sources than the outer regions. Thus, one must take
into account this effect while computing the logN-logS, that is,
the number of sources whose flux exceeds a given flux limit.

As a preliminary step to evaluating the sky coverage, we computed
background maps for the two energy bands using the Ximage package. We
did this by subtracting the central part of the field (the same $\sim
2' \times 2'$ region centered on the 3C 295 cluster excluded from the
source detection) and excluding all the sources detected in the
images. In order to calculate the sky coverage, these maps and the
exposure maps were rebinned into $128 \times 128$ matrices of $\sim
10$ arcsecs per pixel.

We also studied the increasing apparent size of the detected sources
with the off-axis angle (i.e., the distance of the source from the aim
point); both quantities are supplied by the detection algorithm.  We
used the rebinned background maps to calculate the background for all
the points in the field in different size regions according to the
relation between apparent size and off-axis. The local background
level was used to calculate the minimum number of counts needed to
exceed the noise in the case of Poisson statistics, to exceed the
probability threshold of $2\times 10^{-5}$. This value assures a
number of spurious sources per field $<1$ in all energy bands (see
previous subsection).  The minimum counts were divided by the exposure
map and thus converted to minimum detectable fluxes. The sky coverage
at a given flux $f$ is simply the sum of all the regions of the
detector whose minimum detectable flux is lower than $f$, and can be
easily converted into $deg^2$.  The sky coverage computed with such a
method for the three energy bands is shown in fig. 3.

\begin{figure}
\centering
\includegraphics[angle=0,width=8cm]{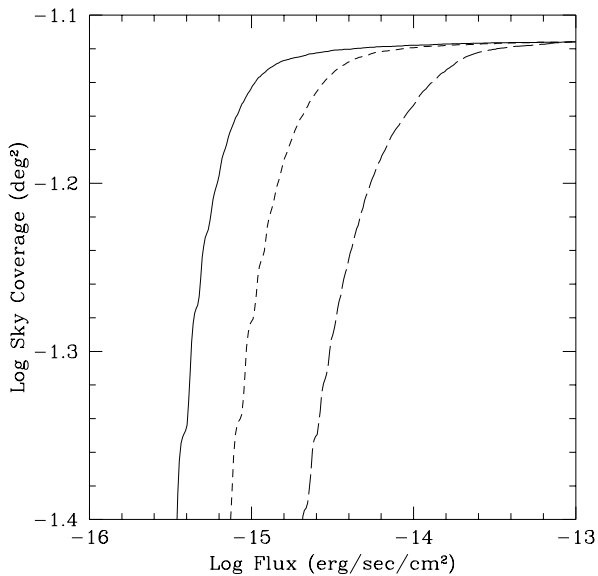}

\caption{The sky coverage (area covered vs. flux limit) for the
$0.5-2$ keV band (solid line), for $2-10$ keV band (long dashed line)
and for the $0.5-10$ keV band (short dashed line).  }
\label{spe1}
\end{figure}

\section{Results}

%
%

\subsection{logN-logS}

\subsubsection{Checking the analysis method}

To check the validity of our analysis method, we computed the
LogN-LogS of the CDFS and we compared this with previous 
computations (Rosati et al. 2002); the agreement is quite good (see
fig. 4). 
Then we computed the logN-logS for 3C 295 in the three energy bands weighting the
number of sources at a given flux with the sky coverage at the same
flux; in fig. 5 is shown the LogN-LogS for the $0.5-2$ and $2-7$ keV
band. The CDFS logN-logS extends to lower fluxes than the 3C 295
one, because the exposure time is about $10$ times longer, and the
error are similar in the region of flux overlap for the two field.
There is excellent agreement between the number counts for the two
fields.

\begin{figure}
\centering
\includegraphics[angle=0,width=9cm]{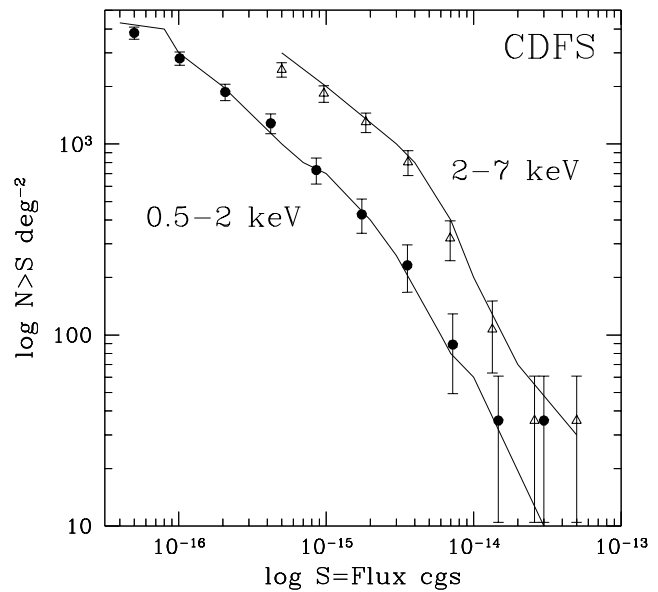}
\caption{The {\it Chandra} Deep Field
South logN-logS in the $0.5-2$ keV band (filled circles)
and in the $2-10$ keV band (open triangles) evaluated in this paper. 
Errors represent $1\sigma$ confidence limit. Solid lines represent the
CDFS LogN-LogS from Rosati et al. (2002).  }
\label{spe1}
\end{figure}

\begin{figure}
\centering
\includegraphics[angle=0,width=9cm]{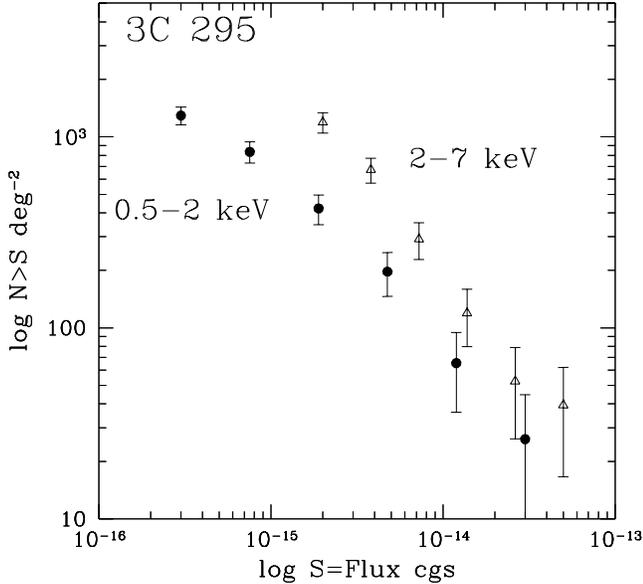}
\caption{The 3C 295 logN-logS in the $0.5-2$ keV band (filled circles)
and in the $2-10$ keV band (open triangles) evaluated in this paper. 
Errors represent $1\sigma$ confidence limit.}
\label{spe1}
\end{figure}

\subsubsection{Chip-by-chip LogN-LogS}

While the average 3C 295 logN-logS appears normal, our goal is to
search for inhomogenities within the ACIS-I field. The most
  natural way to do it
is to derive the logN-logS for each $8' \times 8'$
ACIS-I chip, separately.  The logN-logS for each ACIS-I chip, in each
of the three bands, is shown in Fig. 6. The logN-logS in the upper
left quadrant is significantly higher than that of the lower right
chip, up to $2\times 10^{-15}$ ergs cm$^{-2}$ s$^{-1}$ in the $0.5-2$
keV band, up to $6\times 10^{-15}$ ergs cm$^{-2}$ s$^{-1}$ in the
$2-10$ keV band, and up to $10^{-14}$ ergs cm$^{-2}$ s$^{-1}$ in the
$0.5-10$ keV one.

To be more quantitative, we fitted a power law to the four ACIS-I
logN-logS in each band, in the $3\times 10^{-16} - 2\times 10^{-14}$
ergs cm$^{-2}$ s$^{-1}$ flux range for the soft band, $3\times10^{-15}
- 2\times 10^{-14}$ ergs cm$^{-2}$ s$^{-1}$ for the hard band, and
$6\times10^{-16} - 2\times 10^{-14}$ ergs cm$^{-2}$ s$^{-1}$ for the
$0.5-10$ keV band. Table 1 shows the results of the LogN-LogS fits for
the two fields. In Fig. 7 we plot the slope of the LogN-LogS power law
versus the normalization at $5\times10^{-16}$ ergs cm$^{-2}$ s$^{-1}$
for the soft band, at $5\times10^{-15}$ ergs cm$^{-2}$ s$^{-1}$ for
the hard band and at $10^{-15}$ ergs cm$^{-2}$ s$^{-1}$ for the total
band; error bars represent the $90 \%$ confidence limit. Though the
slopes of the four logN-logS are all compatible within the errors, the
normalization of the NE LogN-LogS chip is clearly higher than that of
the three other chips for all three bands.  The best fit for the CDFS
is shown with dashed lines and appear to be consistent with the latter
chips, both for slope and normalization. To give some numbers, the NE
chip differs in normalization from the SW one by $3.2 \;\sigma$ in the
soft band, by $3.3 \;\sigma$ in the hard one and by $4.0 \;\sigma$ in
the total band. We note that in the total band, the significance of
the discrepancy between the NE and SE chip is higher. This is due to
the higher statistics that can be achieved using the whole energy
range of the observation.

\begin{figure}
\includegraphics[angle=0,width=7cm]{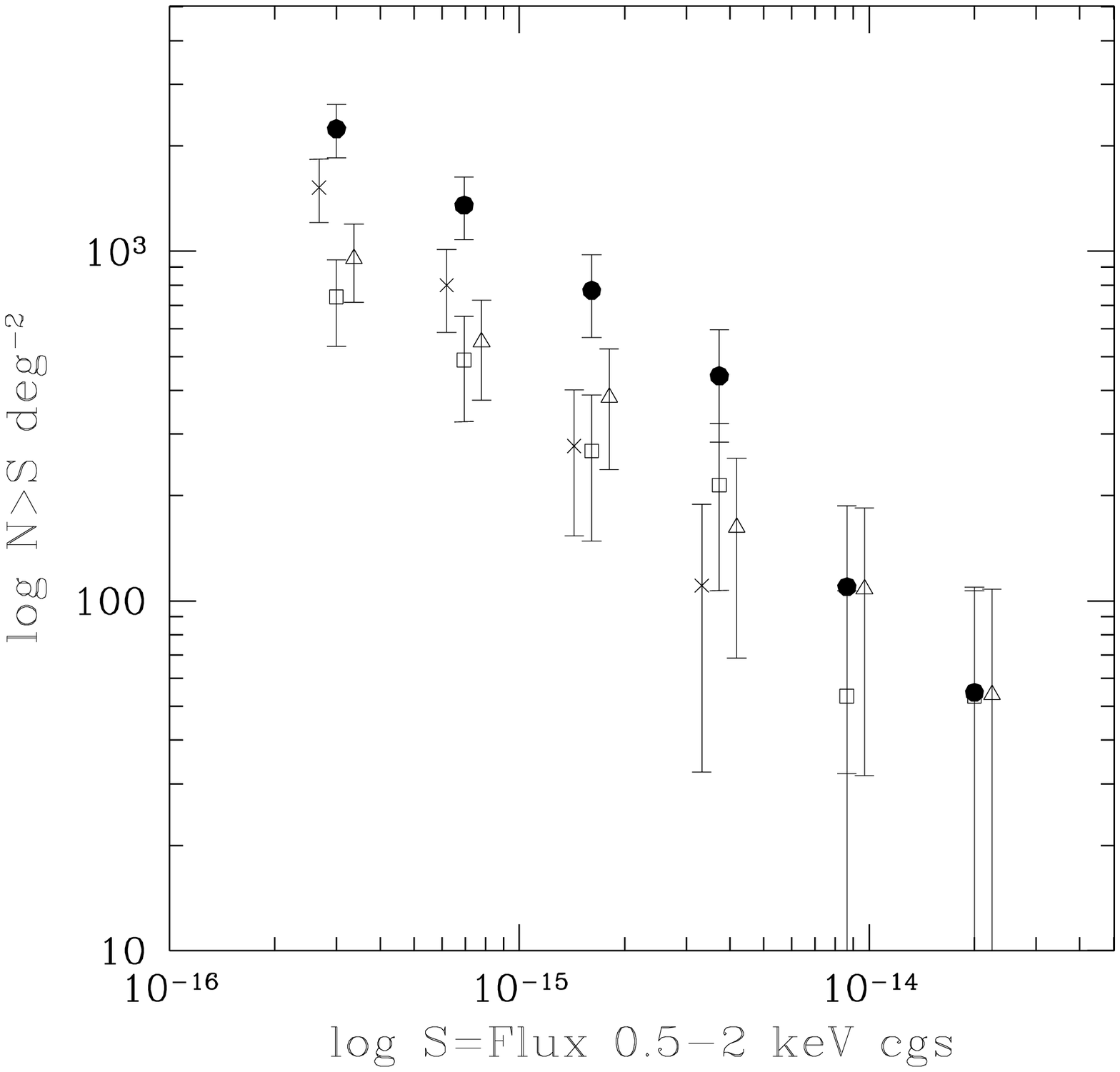}

\includegraphics[angle=0,width=7cm]{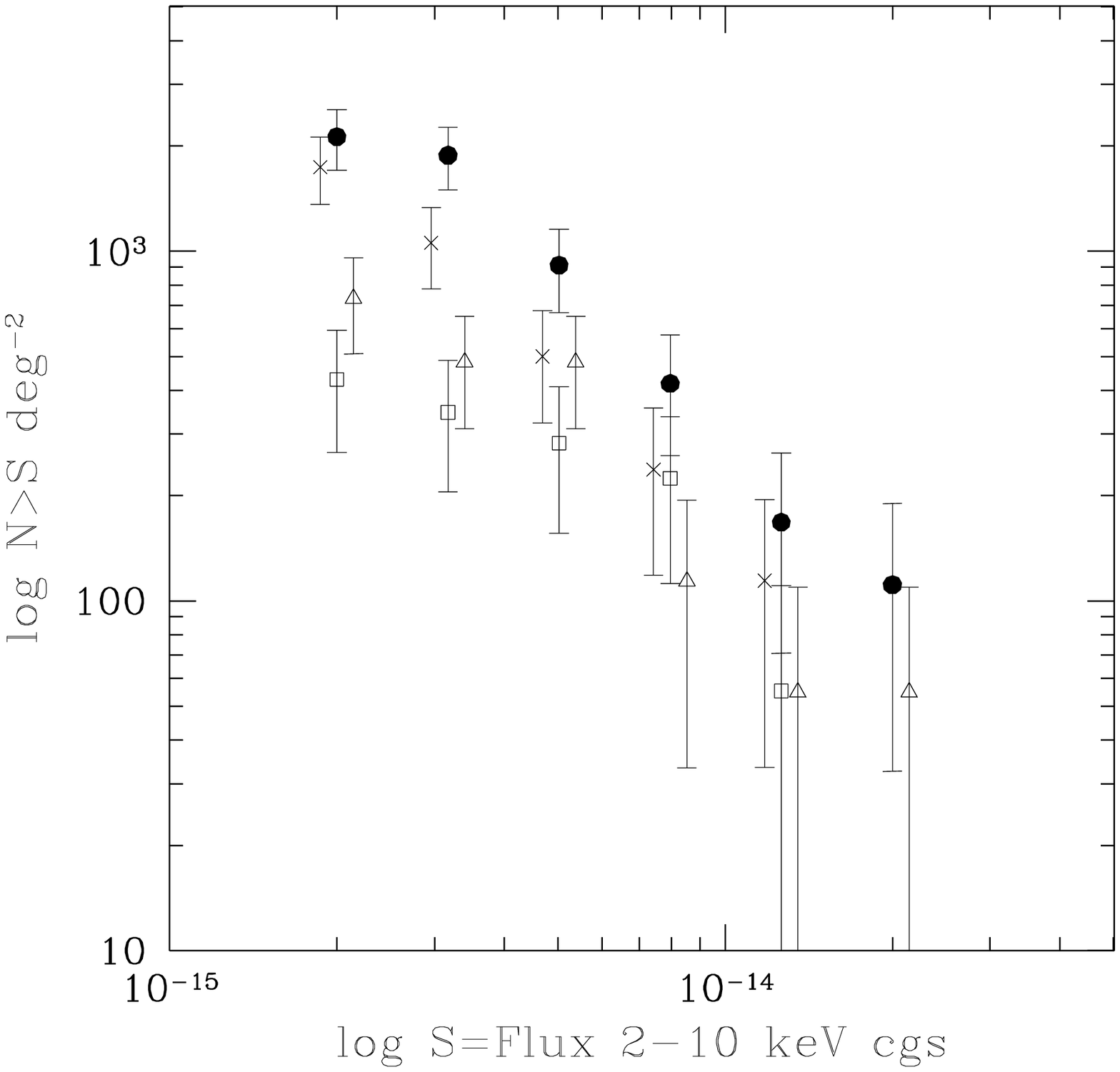}

\includegraphics[angle=0,width=7cm]{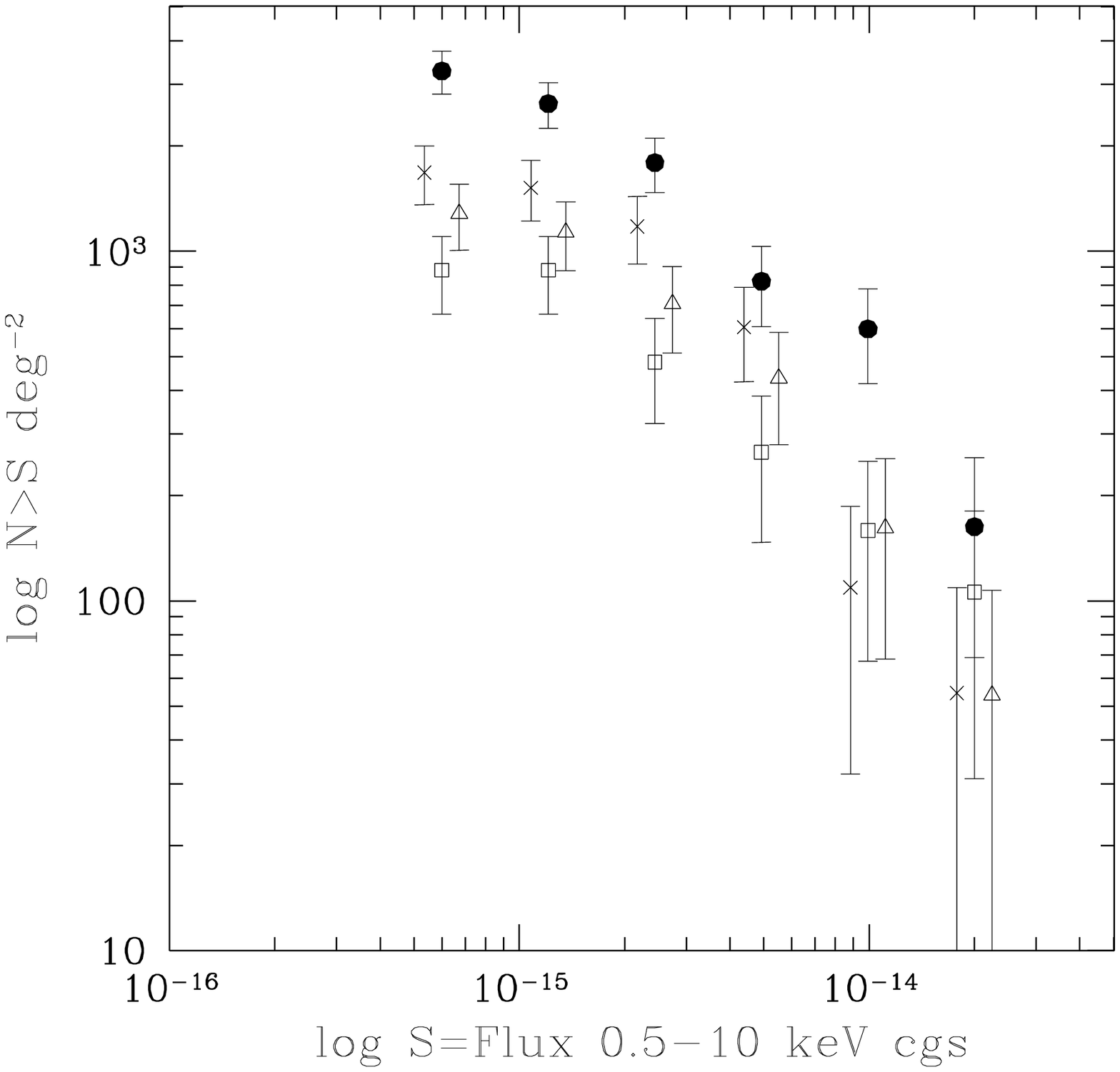}
\caption{ The mean (whole field) 3C 295 logN-logS in the $0.5-2$ keV
band (upper panel), in the $2-10$ KeV band (central panel) and in the
$0.5-10$ keV band (lower panel), calculated for each ACIS-I chip
separately. Filled circles represent counts for the NE chip, open
triangles for the NW, open squares for the SW and crosses for the SE.
For clarity reasons, the points for SE chip and the NW have been
shifted slightly to the left and to the right, respectively. Errors
represent $1\sigma$ confidence limit.  }
\label{spe1}
\end{figure}

\begin{table*}[ht]
\caption{\bf Results of the four 3C 295 chips and CDFS LogN-LogS fit }
\footnotesize
\smallskip
\begin{tabular}{|l|c|c|c|c|c|c|c|c|c|}
\hline
\multicolumn{1}{|l|}{}&
\multicolumn{3}{|c|}{$0.5-2$ keV}&
\multicolumn{3}{|c|}{$2-10$ keV}&
\multicolumn{3}{|c|}{$0.5-10$ keV} \\
\hline
 & sources & slope 	& normaliz. & sources
&slope&normaliz.& sources& slope&normaliz.\\ 
\hline
3C 295  & $89$& $-0.75^{+0.09}_{-0.1}$       &$1000^{+100}_{-100}$&$71$&$-1.3^{+0.2}_{-0.2}$&$390^{+60}_{-50}$&$ 121$&$-0.9^{+0.1}_{-0.2}$&$ 1300^{+200}_{-200}$\\
\hline
NE	& $36$& $-0.8^{+0.1}_{-0.2}$& $1600^{+600}_{-400}$&$29$&$-1.6^{+0.4}_{-0.6}$&$900^{+300}_{-300}$ &$53$&$-0.7^{+0.1}_{-0.1}$ & $ 2600^{+600}_{-600}$\\
\hline
NW	&$18$& $-0.7^{+0.2}_{-0.3}$	& $700^{+300}_{-400}$&$13$& $-1.4^{+0.7}_{-0.9}$&$300^{+200}_{-100}$ &$23$&$-0.7^{+0.2}_{-0.2}$ & $1000^{+300}_{-300}$\\
\hline
SW	&$12$& $-0.7^{+0.2}_{-0.3}$	& $600^{+300}_{-300}$&$7$&$-1.1^{+0.7}_{-1.0}$&$200^{+200}_{-100}$ &$17$&$-0.6^{+0.2}_{-0.2}$ & $700^{+400}_{-300}$\\
\hline
SE	&$23$& $-1.0^{+0.3}_{-0.4}$	& $950^{+400}_{-300}$&$22$&$-1.6^{+0.7}_{-0.9}$&$ 500^{+200}_{-200}$ &$28$&$-0.8^{+0.1}_{-0.2}$ & $1300^{+500}_{-300}$\\
\hline
CDFS	&$202$& $-0.9^{+0.1}_{-0.1}$	& $900^{+100}_{-100}$&$151$&$-1.7^{+0.2}_{-0.2}$&$280^{+40}_{-40}$ &- & -&- \\
\hline
\end{tabular}

\normalsize
\end{table*}

\begin{figure}
\centering
\includegraphics[angle=0,width=7cm]{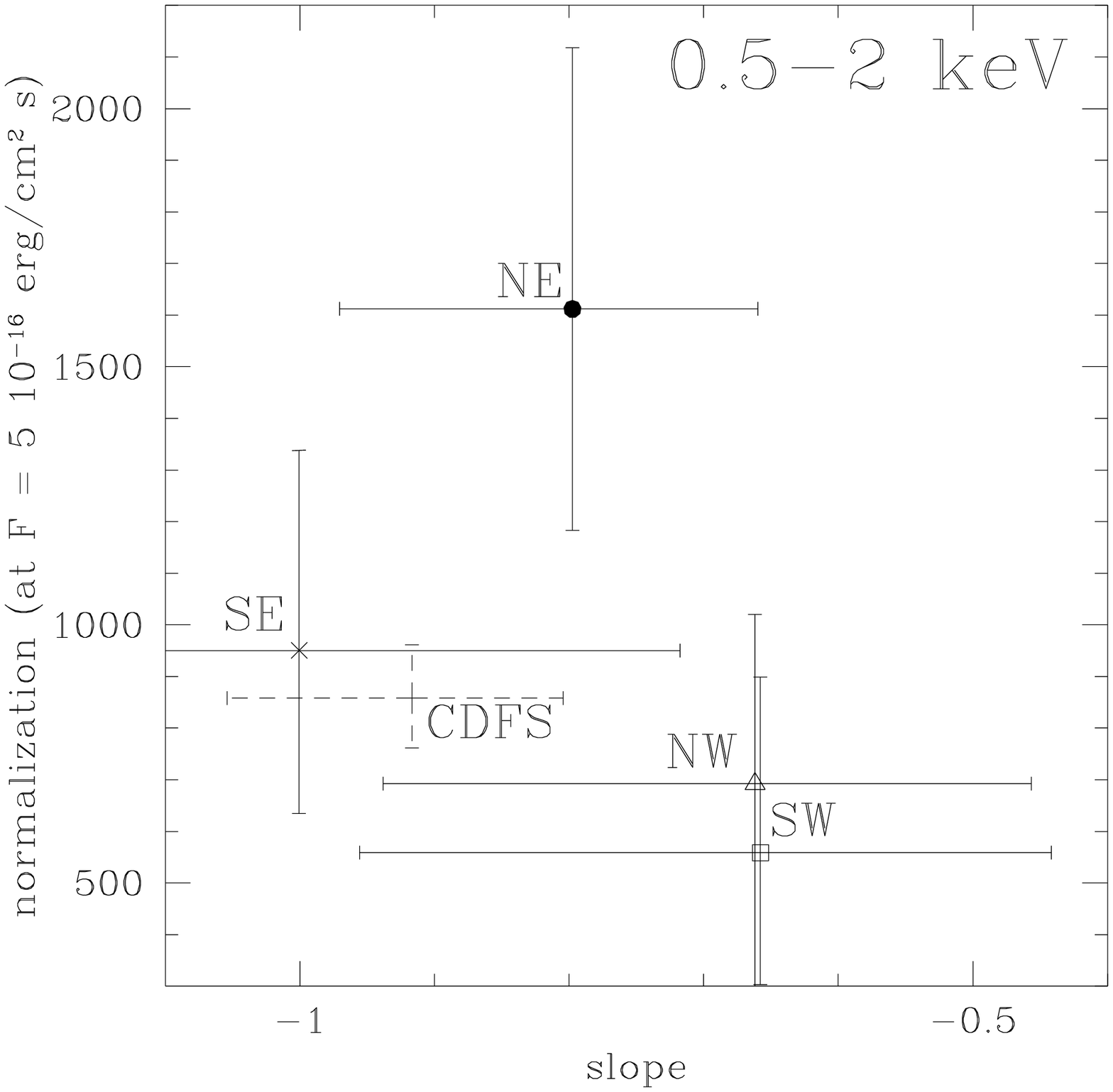}

\includegraphics[angle=0,width=7cm]{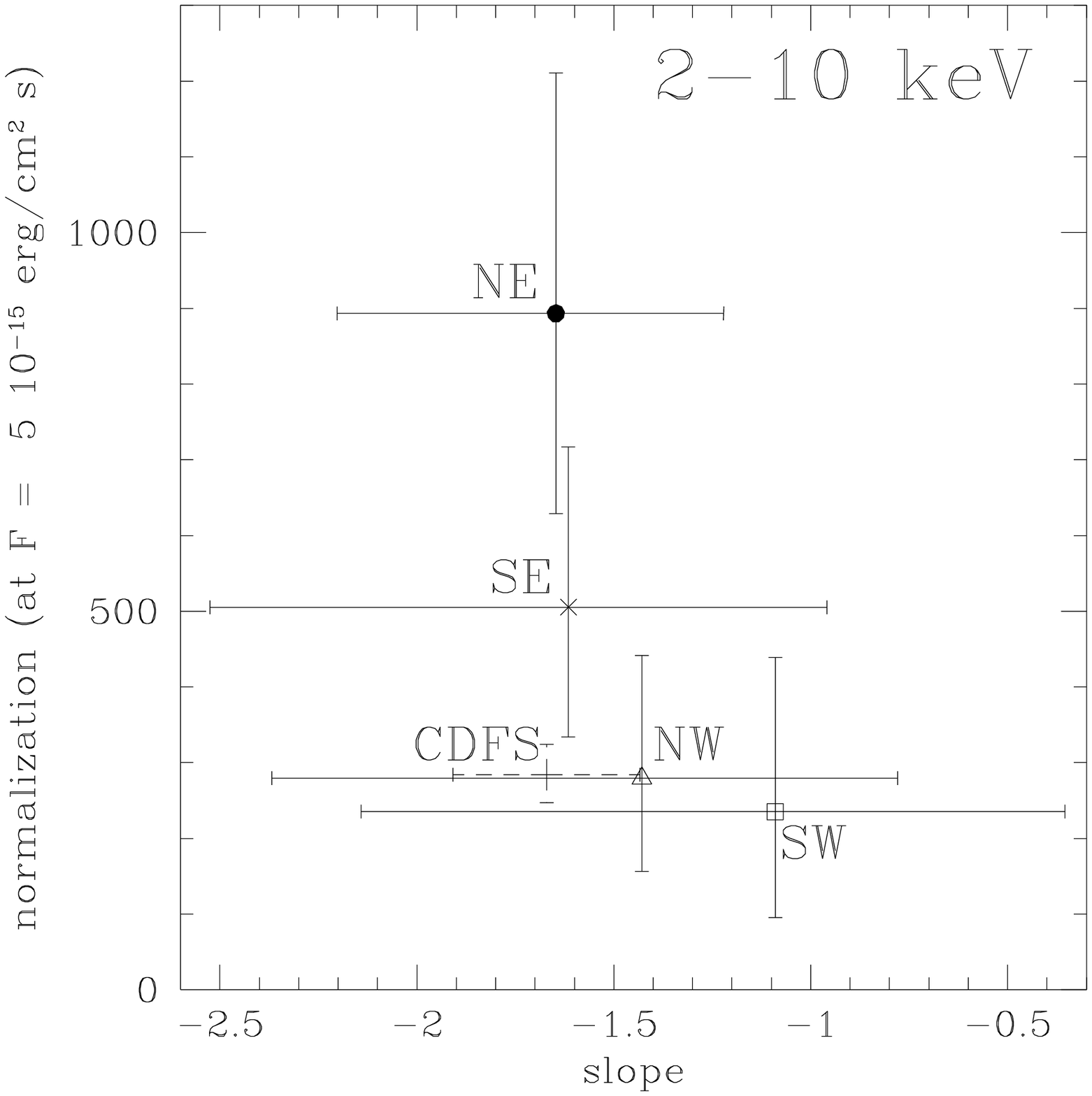}

\includegraphics[angle=0,width=7cm]{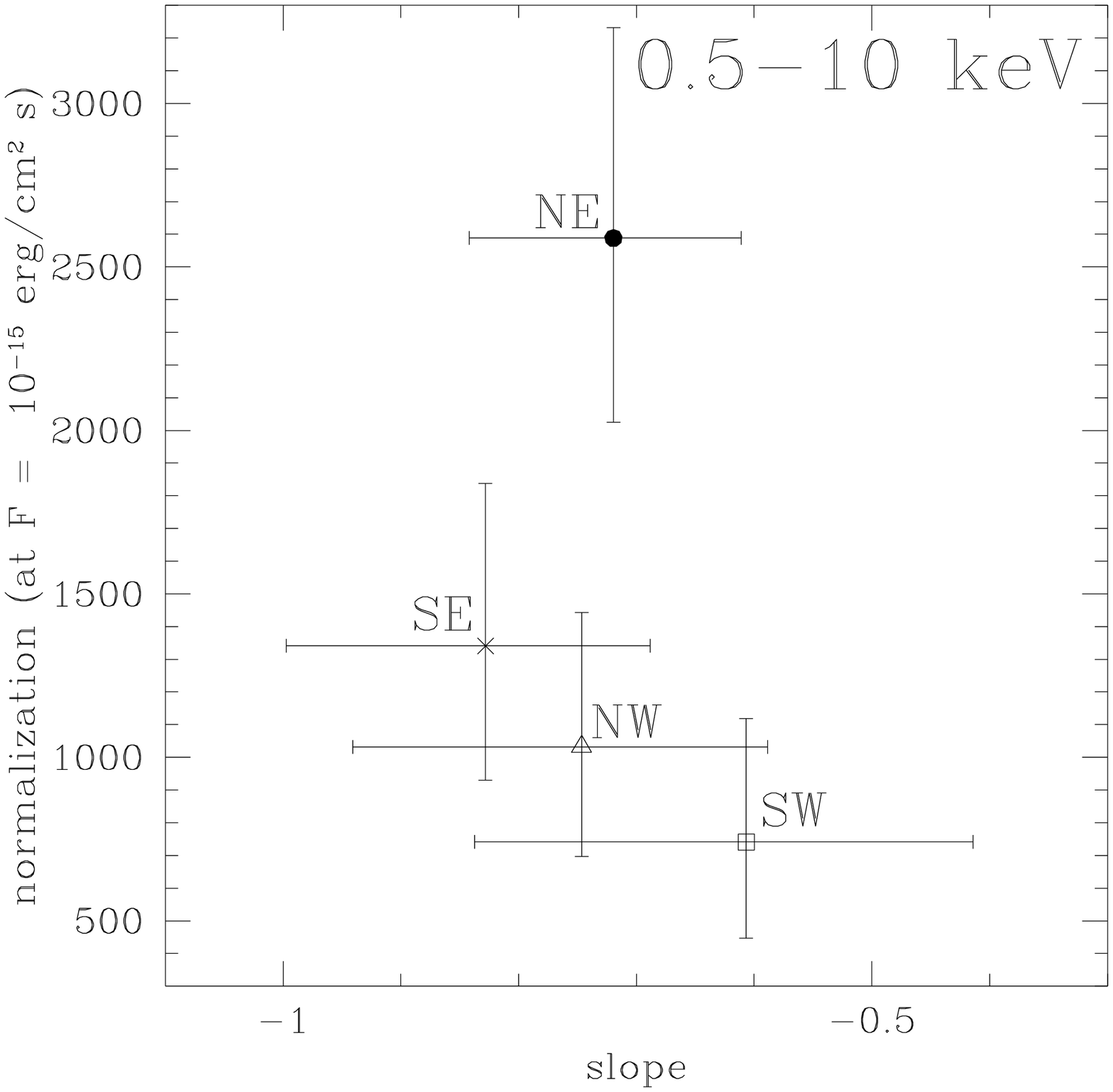}
\caption{ Results of the power law fits to the four logN-logS chips in
the soft band (upper panel), in the hard band (central panel) and for the
whole band (lower panel). x axis plots the slope of the power law,
while on the y axis is plotted the normalization at $5\times 10^{-16}$
ergs cm$^{-2}$ s$^{-1}$ (soft band), at $5\times 10^{-15}$ ergs
cm$^{-2}$ s$^{-1}$ (hard band) and at $ 10^{-16}$ ergs cm$^{-2}$
s$^{-1}$ (whole band).  Filled circles represent the NE chip, open
triangles the NW, open squares the SW and crosses the SE. Dashed lines
represent the fit for the CDFS. Errors are the $90\%$ confidence
limit.  }
\label{spe1}
\end{figure}

In the next subsections we will study in more detail the distribution
of the sources, in particular how much it differs from the uniform
distribution, and its angular correlation function.

\subsection{Two-dimensional Kolmogorov-Smirnov test}

In this subsection we apply a two-dimensional Kolmogorov-Smirnov (KS)
test to check whether the distribution of the sources is uniform or
not.  The 1-D KS test is a simple and powerful tool to test whether
two data sets are drawn from the same distribution, or whether a data
set is compatible with a given distribution, provided that these data
sets and distributions are functions of only one variable. The KS test
makes use of the parameter $D$, which represents the maximum value of
the absolute difference between the cumulative distributions drawn
from the two data sets, or from the data set and the given
distribution. Once $D$ is known, the probability that a larger value
of $D$ can be observed if the two distributions are identical is given
by:
$$ P(D>observed) = 2\,\sum_{i=1}^{\infty} (-1)^{i-1}e^{-2\,i^2\,\lambda^2},
\eqno (1)$$  
where $\lambda$ is a simple function of $D$.

The generalization of the KS test to a two-dimensional distribution is
due to Peacock (1983) and Fasano \& Franceschini (1987). Any point of
a data set is now defined by a pair of coordinates, which in turn
define a plane. Thus, given a point of the data set, one can compute
how many elements (of the same data set or of another one) occupy the
four quadrants defined by such a point, or how many points are
expected in the four quadrants according to the given
distribution. The parameter $D$ can now be defined as the maximum
value of the absolute difference between the number of elements of the
two data set (or of the data set and the distribution) in the
quadrants, as the reference point changes within the data set.  Fasano
\& Franceschini (1987) demonstrated that eq. (1) holds even for
two-dimensional distributions, provided that
$$ \lambda = {\sqrt{N}\,D\over 1 + \sqrt{1-r^2}\;(0.25-0.75/\sqrt{N})};
\eqno (2)$$
here $N$ is the number of elements of the data set (or
$N=N_1\,N_2/(N_1+N_2)$ if two data sets are involved) and r is the
correlation coefficient of the two variables of the data set (or the
average between the two correlation coefficients in the two data set
case).

We have used eq. (1) and (2) to test the hypothesis that the
distribution of the sources in the 3C 295 field is uniform.  Since the
sources are not uniformly distributed due to the sensitivity of the
instrument being higher in the middle of the detector (see section
2.2), thus low flux sources can only be found near the centre of the
four ACIS-I chips, so we imposed a higher minimum flux. We specified
this minimum flux as the one which could have been detected at least
in $75\%$ of the detector area. We note that this effect is not very
strong in the 3C 295 field, because the
central part of the detector was excluded due to the presence of the
cluster. We found the probability that the distribution of
the 3C 295 sources is uniform is very low: $\sim 3\%$, $\sim 4\%$
and $\sim 0.6\%$ in the $0.5-2$, $2-7$ and $0.5-7$ keV bands,
respectively (see also table 2).




\subsection{Angular correlation function}

Although the LogN-LogS and the 2-D KS test show that the sources in
the 3C 295 field are not uniformily distributed, they do not give
angular information on the spatial distribution of the sources.  The
conventional way to determine the spatial correlation of sources in a
field is to use the `two point angular correlation function' (ACF,
Peebles 1980, Peacock 1999). This function measures the strength of
the spatial correlation, but does not say anything about where the
sources are clustered. This function is given by the following
relation:
$$w(\theta) = {DD\over RR} -1, \eqno (3)$$ 
where $DD$ is the ratio of pairs of sources in the real data, which
fall in the separation angle ($\theta,\, \theta + d\theta$) and the
$RR$ is the number of those pairs expected when the spatial
distribution of sources is random. To compute the ACFs, we proceeded
as follows: we computed the angular separation for the pairs in our
data set; we generated a random uniform distribution of $1000$ sources
in the same area and we computed their angular separation; we
evaluated the ACFs using eq. (3) and rescaling $RR$ to the number of
pairs of the real field.  We made the same flux cut-off to take into
account the higher sensitivity of the detector in its central region
(see previous subsection). Fig. 8 shows the ACFs for the 3C 295 field.

\begin{figure}
\centering 
\includegraphics[angle=0,width=7cm]{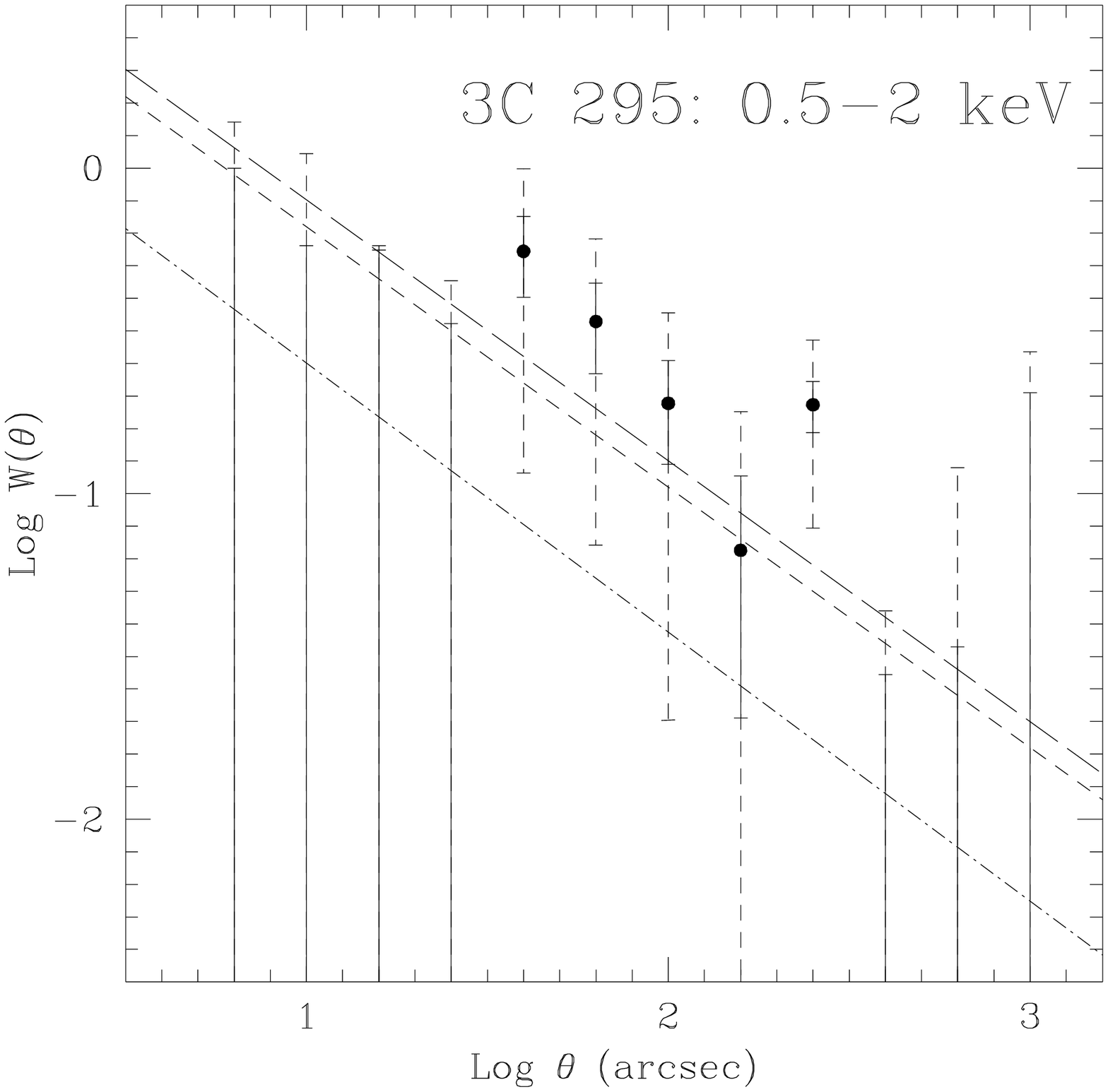}

\includegraphics[angle=0,width=7cm]{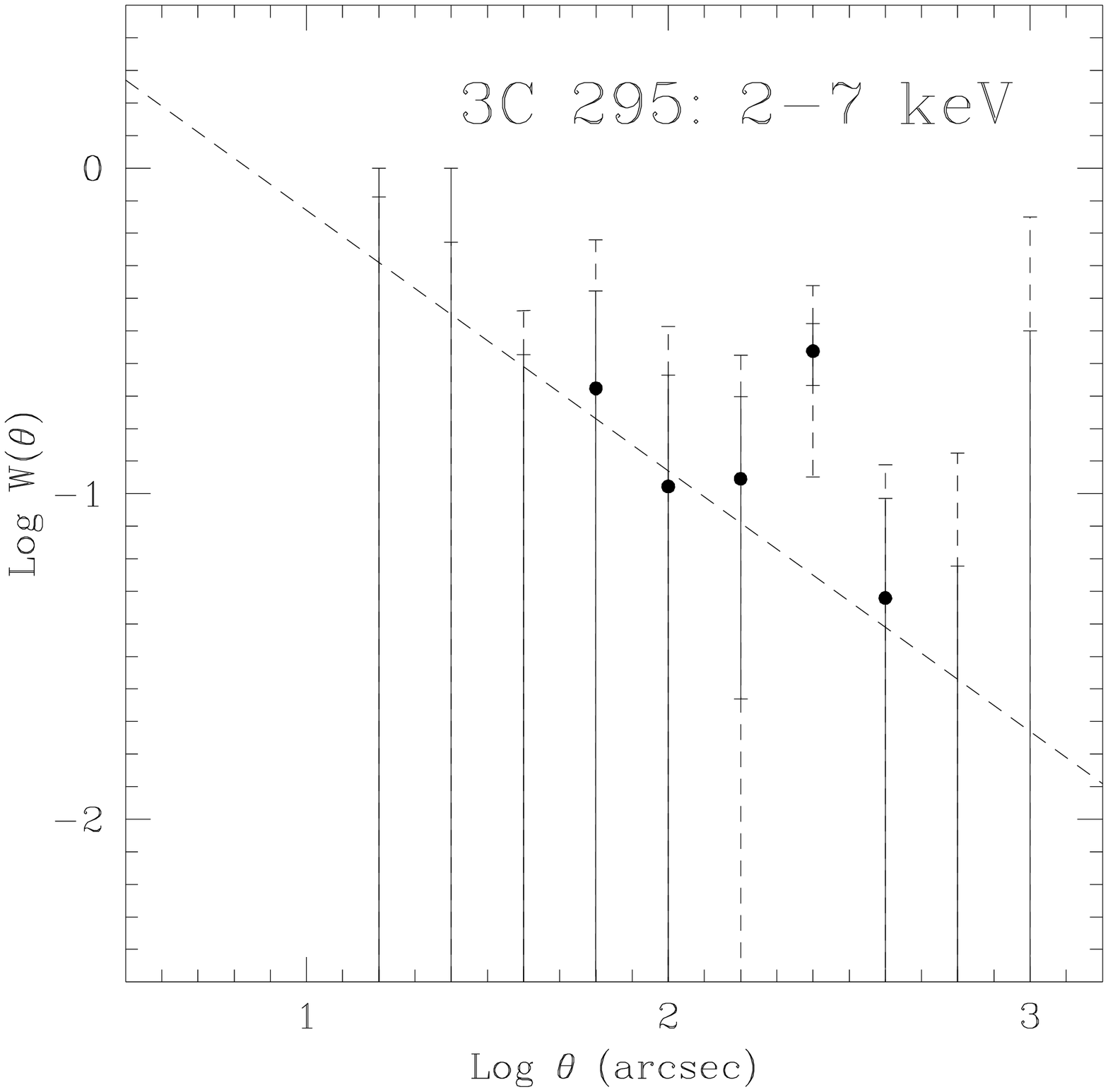}

\includegraphics[angle=0,width=7cm]{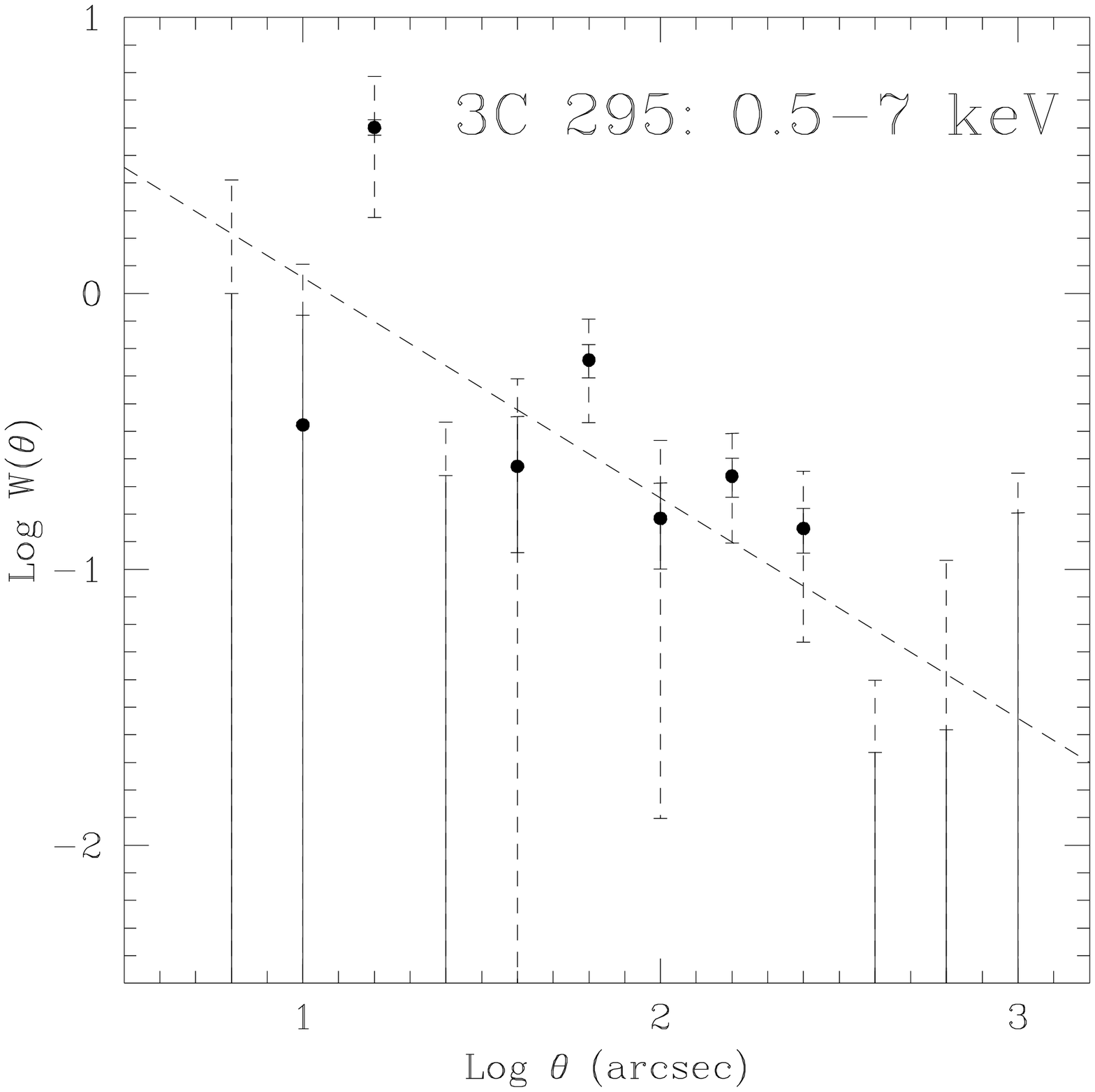}
\caption{ The 3C 295 angular correlation function in the $0.5-2$ keV band
(upper panel), in the $2-7$ keV (central panel) and in the $0.5-7$ keV
  band (lower panel). Solid error bars are Poisson;
dashed error bars are bootstrap. Short-dashed lines represent our fit to
  the data assuming bootstrap error bars. 
The long dashed line represents the best fit found by 
Vikhlinin \& Forman (1995) for ROSAT sources; the dot-dashed
line is the simulated ACF calculated by the same authors in order to
correct the effect of the amplification bias.
}
\label{spe1}
\end{figure}



The errors in fig. 8 were estimated using two different methods.
The first one (solid lines) is simply the Poisson statistics, the
error is defined as the inverse of the square root of the real pair
number (Peebles 1980) and is show in fig. 8  with solid lines.
The second one (dashed lines) is more accurate  and is called the
bootstrap resampling method (see Barrow, Bhavsar \& Sonoda 1984 and
references therein). Briefly, we generated $100$ simulated fields from
the real one, by randomly selecting within the sources of the field a
number of sources equal to the data set size.  This produces simulated
fields with a number of elements equal to that of the real field, but
with some sources counted more than once and others not considered at
all. The bootstrap method consists in recalculating the ACF using such
simulated fields and in calculating the error as the standard
deviation of the $100$ ACFs obtained in this way.  The bootstrap
errors (which probably overestimate the true error) 
exceed the Poisson ones. This is because the Poisson errors
depend on the number of pairs at each scale, while the bootstrap ones
also depend on how many pairs are produced by each single source. In
more detail, let us suppose that a large number of sources are
distributed along a circumference, and one is placed at its centre;
obviously, a strong angular correlation will be found at scales equal
to the circle's radius. Poisson errors will be strongly reduced as the
number of sources along the circumference increases, ignoring that a
single source (namely, the central one) strongly boosts the
correlation; bootstrap errors take this effect inThe other way roundto account by
randomly excluding or counting more than once the central source, when
recomputing the ACF.
  
Fig. 8 (long dashed line) also shows the best fit for the angular correlation
function found by Vikhlinin \& Forman (1995) for an extensive set of
deep ROSAT observations covering $ 40$ deg$^2$ of sky. Their power law
model $w(\theta) = (\theta/\theta_0)^{(1-\gamma)}$ yields the best fit
parameters $\theta_0 = 10\pm 8$ arcsecs and $\gamma = 1.7 \pm
0.3$. 
However they found that the correlation angle
$\theta_0$ was smaller than the FWHM of the ROSAT PSPC PSF, which is
$\sim 25$ arcsec on-axis, which leads to an amplification bias. 
Simulations to correct for
this gave the ACF show in Fig. 8 with a dot-dashed line, which
has a normalization 2.85 times smaller (Vikhlinin \& Forman 1995). 
This means that the 
uncorrected Vikhlinin \& Forman (1995) ACF (long dashed line) 
represents a strong upper
limit to the ROSAT ACF. Since the ROSAT PSPC energy band is 0.1-2 keV
we cannot compare the VK85 ACF with our ACFs in the hard (2 - 7 keV) and 
whole (0.5 - 7 keV) bands.


Vikhlinin \& Forman (1995) sample could only
produce the ACF down to a $\sim 30$ arcsec scale ($Log \theta =
1.5$). 

For the 3C 295 field in the $0.5-2$ keV band, we found a signal above
the 
upper limit
of the Vikhlinin \& Forman (1995) fit on scales 
$\sim 60\;-\;240$ arcsec,
if we use the Poissonian error bars; however using the bootstrap
errors reduces this scale to only $\sim 240$ arcsec. A similar effect
is present in the $2-7$ keV band, where 
a clear signal of  correlation emerges
on scales of $\sim 240$ arcsec assuming both Poisson and bootstrap
statistics.  A stronger and wider signal is present in the $0.5-7$ keV band,
on scales ranging from $\sim 15 $
arcsec to $\sim 240$ arcsec.  This reflects the
clustering of the sources clearly visible in the upper left corner of
fig. 1 (on linear scales of the order of half a chip length, $\sim
240$ arcsec).  The improved statistics in the $0.5-7$ keV band, due to
the higher number of sources detected, gives rise to a stronger signal.

The short-dashed lines in fig. 8 represent the fit 
of $w(\theta) = 
(\theta/\theta_0)^{1-\gamma}$ to our data,
assuming the bootstrap error bars; $\gamma = 1.8 $ was fixed as
in Vikhlinin \& Forman (1995).
In the soft band we found $\theta_0=6.0^{+4.4}_{-3.7}$ arcsec, in the hard
band $\theta_0=6.9^{+6.9}_{-5.1}$ arcsec, and in the broad band 
$\theta_0=8.5^{+6.5}_{-4.5}$ arcsec ($90$\% confidence limit). These values are consistent
within their relatively large error to the
 Vikhlinin \& Forman (1995) fit for $\theta_0$ without correcting for the
 amplification bias ($\theta_0 \sim 10$ arcsec). 
If we assume the Poisson error bars, the value of $\theta_0$ rises and
the errors become smaller in all energy bands. 

Finally, we checked the dependance of the ACF on the flux limit. We computed
the ACF in the broad band for three different flux limits (i.e., $1$
times, $2$ times and $3$ times the flux limit given in tab. 2,
footnote $a$) and we found
that $\theta_0$ vary less than $25\%$, while its error increases
with the flux limit.

\subsubsection{CDFS}

We applied the KS test and evaluated the ACF for the CDFS
in order to check whether this field is different from 3C
295.

Table 2 shows the results of the two dimensional KS test. We can see that 
CDFS is compatible with a uniform distribution with a higher 
probability ($\sim 15 \%$) than 3c 295 ($\sim$ a few $\%$). 
We note that the KS test result depends on the minimum flux chosen; 
nevertheless, KS for CDFS gives a higher probability of a 
uniform distribution than 3C295 even if we use for the two
observations the same flux limit. The first and the third column of
table 2 report the KS results for the flux limit described in this 
subsection; the second column shows the CDFS results for a
flux limit equal to that of 3C 295.

In fig. 9 we show the ACF for the CDFS field; in both the energy bands we found
no evidence of a strong signal such as that featured by the 3C 295
field. Giacconi et
al. (2001) evaluated the CDFS ACF for the first 100 ks observation of
the CDFS; so, we checked our algorithm by calculating the ACF for the
same data and found, reassuringly, agreement within $1\;\sigma$.

\begin{table*}[ht]
\caption{\bf KS test for CDFS and 3C 295 fields}
\footnotesize
\smallskip
\begin{tabular}{|l|c|c|c|}

\hline
 & 3C 295$^a$ 	& CDFS$^a$ & CDFS$^b$	\\ 
\hline
$0.5-2$ keV	& $3.09\;10^{-2}$&$0.2$	& $0.13$\\
\hline
$2-7$ keV	& $3.84\;10^{-2}$&$0.5$	& $0.17$\\
\hline
$0.5-7$ keV	& $6.32\;10^{-3}$	& -&- \\
\hline
\end{tabular}

\medskip
$^a$ lux limits: $0.5-2$ keV, $F_X > 4.5\times 10^{-16}$ ergs
s$^{-1}$ cm$^{-2}$; $2-7$ keV, $F_X > 3.5\times 10^{-15}$ ergs
s$^{-1}$ cm$^{-2}$; $0.5-7$ keV, $F_X > 7.8\times 10^{-16}$ ergs
s$^{-1}$ cm$^{-2}$. 

$^b$ flux limits: $0.5-2$ keV, $F_X > 6.0\times 10^{-17}$ ergs
s$^{-1}$ cm$^{-2}$ $2-7$ keV, $F_X > 5.0\times 10^{-16}$ ergs
s$^{-1}$ cm$^{-2}$.

\normalsize
\end{table*}

\begin{figure}
\centering 
\includegraphics[angle=0,width=8cm]{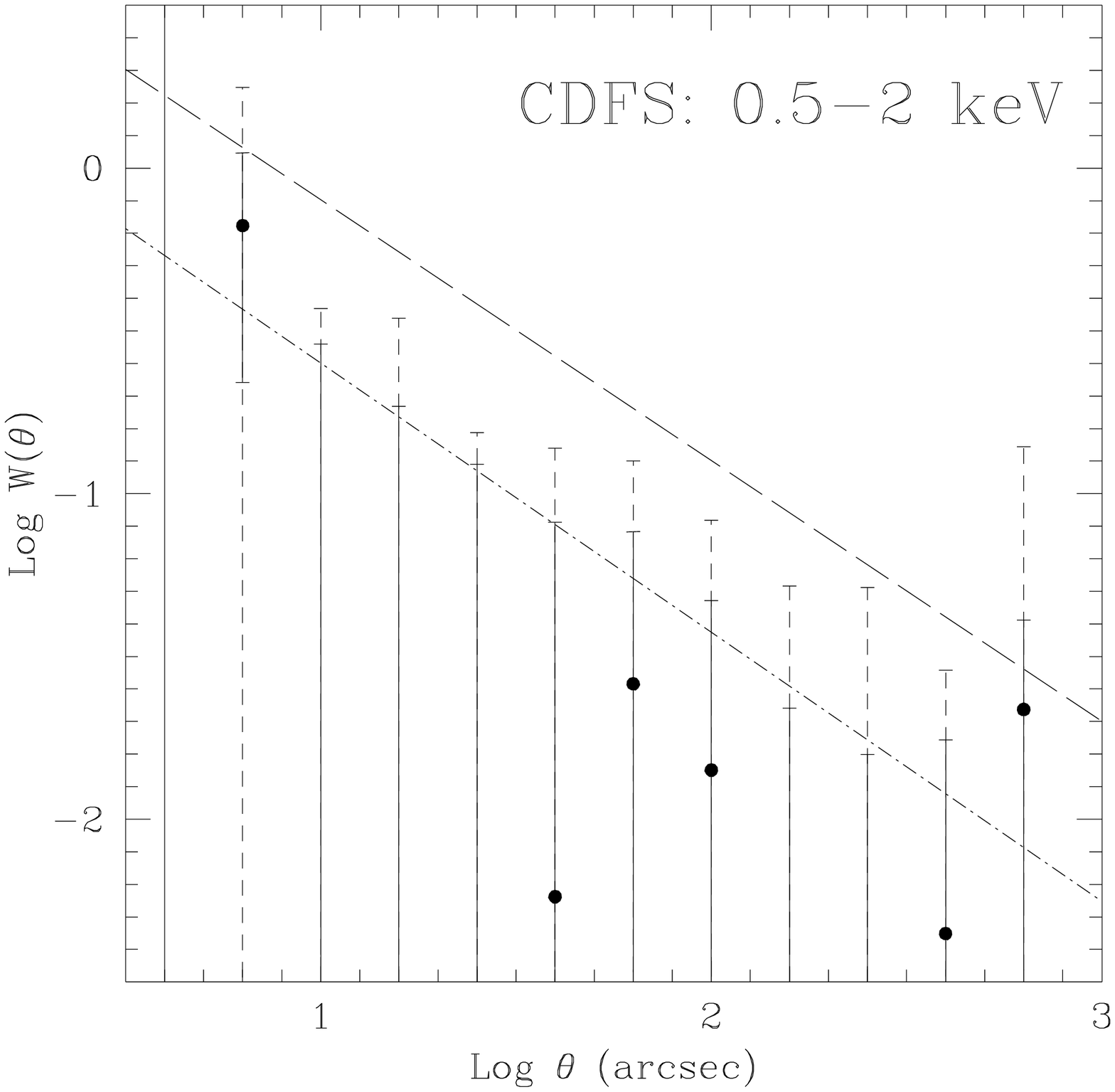}

\includegraphics[angle=0,width=8cm]{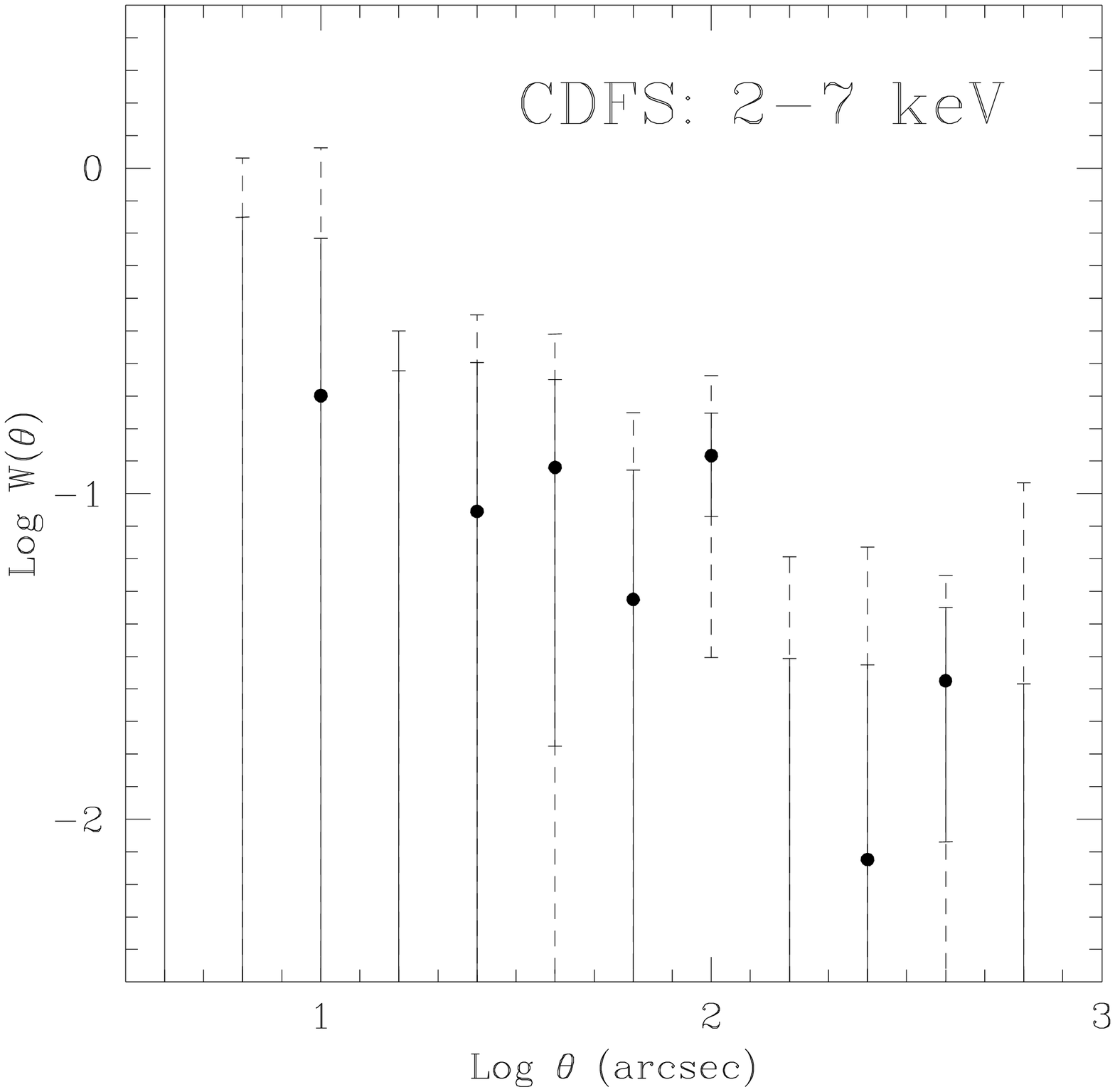}
\caption{The CDFS angular correlation function in the $0.5-2$ keV band
(upper panel) and in the $2-7$ keV  (lower panel). The dashed line  
line represents the best fit found by 
Vikhlinin \& Forman (1995) for ROSAT sources; the dot-dashed
  line is the simulated ACF calculated by the same authors in order to
  correct the effect of the amplification bias. Solid error bars are Poisson;
dashed error bars are bootstrap.
}
\label{spe1}
\end{figure}

\subsubsection{ACF in 8 selected {\it Chandra} Fields}

In order to check whether or not the strong clustering of the 3C 295 field is peculiar,
the comparison with the CDFS is not enough, and a larger sample of {\it Chandra} ACIS-I
fields of similar exposure time has to be analyzed. For this reason, we browsed the {\it Chandra}
data archive and matched all the ACIS-I observations belonging to the classes ''AGNs'' and 
''Surveys'', with an exposure time of at least 50 ks. We used the detection algorithm PWDetect
on these fields to detect sources in the three energy bands.
In tab. 4 we show the $8$ fields selected for this analysis, their exposure times and
the sources detected in the three bands.

\begin{table*}[ht]
\caption{\bf {\it Chandra} fields chosen for the ACF analysis}
\footnotesize
\smallskip
\begin{tabular}{|l|c|c|c|c|}

\hline
\multicolumn{1}{|l|}{field}&
\multicolumn{1}{|c|}{exposure time}&
\multicolumn{3}{|c|}{number of sources}\\
 &(ks) & $0.5-2$ keV &$2-7$ keV &$0.5-7$ keV\\
\hline
NGC55 &  60 & 108 & 75 & 132 \\ 
\hline
WHDF & 72 & 83 & 58 & 104 \\
\hline
ISO & 74 & 92 & 76 & 116 \\
\hline
GROTH-WEST.F & 85 & 84 & 63 & 103 \\
\hline
1156+295 & 75 & 145 & 50 & 150 \\
\hline
Elais:N1 & 74 & 101 & 80 & 127 \\
\hline
Elais:N2 & 74 & 85 & 60 & 102 \\
\hline 
FIELD-142549+35 & 122 & 110 & 96 & 148 \\
\hline
\end{tabular}
\normalsize
\end{table*}

We computed the ACF in the $0.5 - 7$ keV band using a flux cut-off
such that the minimum flux could have been detected at least
in $75\%$ of the detector area, as we did for 3C 295. The 
average ACF of this sample is shown in fig. 10. Though a signal is still 
present at scales $> 100 $ arcsec, it is
clearly weaker than that of 3C 295.
To be more quantitative, we fitted these data with a power law of fixed 
slope $1-\gamma$, $\gamma=1.8$ as we did for the 3C 295 ACF, and  we found  
the correlation angle $\theta_0=1.2^{+1.3}_{-1.0}$ arcsec ($90$\% confidence limit). This
is lower that of 3C 295 by a factor of $\gs2$ at a confidence level of $2.5 \;\sigma$.

\begin{figure}
\centering 
\includegraphics[angle=0,width=8cm]{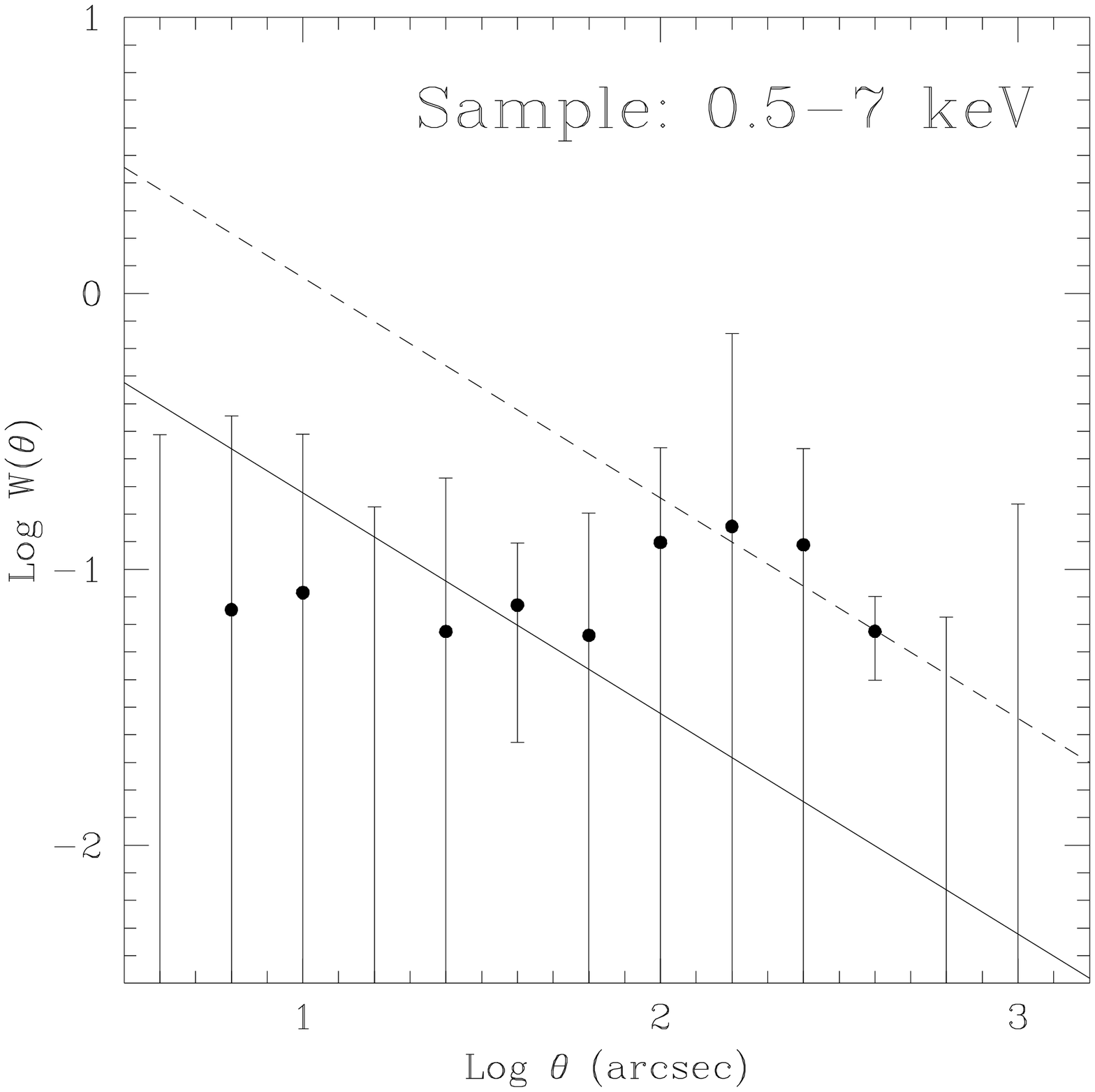}
\caption{ The average angular correlation function in the  $0.5-7$ keV
 band for the sample of fields reported in table 3; Error bars are bootstraps. 
The solid line represents the best fit for these data; the dashed line is the 
fit for the ACF of 3C 295
(see also fig. 8, bottom panel).
}
\label{spe1}
\end{figure}







\section{Discussion}

A $92$ ks {\it Chandra} observation of the source field around the $z
=0.46 $ 3C 295 cluster shows an excess of sources visible in
the NE corner of the {\it Chandra} observation (fig. 1). This is clear
in the density contour plot of fig. 11. This figure shows that the denser 
region of the putative filament has a sources density more than 
$5$ times higher than the average source density of the field ($\sim 0.5$ 
sources per arcmin$^2$. A chip by
chip logN-logS analysis demonstrates that the NE chip has a $4.0
\sigma$ excess of sources over the SW chip in the total ($0.5-7$ keV)
{\it Chanda} band. [In the standard soft ($0.5-2$ keV) band the excess
is $3.2 \sigma$, and in the standard hard ($2 - 7$ keV) band the
excess is $3.3 \sigma$.]  This result confirms the basic result of
Cappi et al. (2001) and extends it to deeper fluxes, larger
field-of-view and the $2-10$ keV band.

\begin{figure}
\centering 
\includegraphics[angle=0,width=9cm]{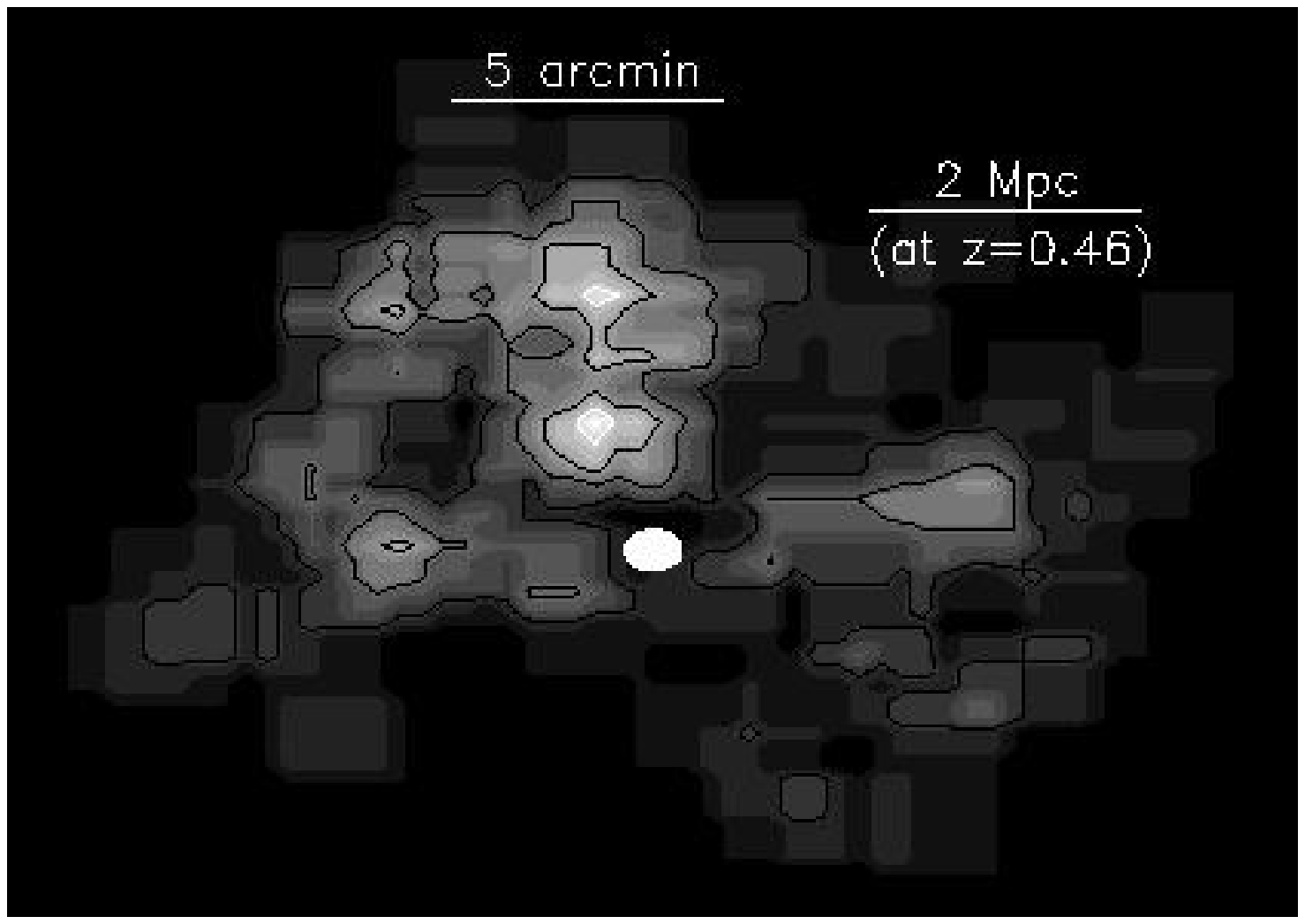}
\caption{The density profile of the 3C 295 field, computed for
the whole $0.5 - 7$ keV band. The linear smoothing factor is $1.5$ arcmin, 
and the four contour levels indicate source densities of $1.3$, $1.8$, $2.2$
and $2.7$ sources per arcmin$^2$. 
The clustering of sources in the NE
corner is clearly visible; the bright central spot is the 3C 295 cluster.
}
\label{spe1}
\end{figure}



The asymmetric distribution of the 3C 295 field is confirmed by two
analyses: (1) the two dimensional Kolmogorov-Smirnov test, which show
the probability that the sources are uniformly distributed is $0.632
\%$ in the total {\it Chandra} band [ $3.09\,\%$ for the soft band and
$3.84\;\%$ for the hard band).  (2) the two point angular correlation
function (ACF) strongly indicate positive correlation, for scales of 
$0.5-5$ arcmins. This strong correlation  has not been found in a sample
of eight ACIS-I fields with a similar exposure time, for which the 
correlation angle $\theta_0$ is smaller by a factor of $\gs2$ 
than that of 3C 295 at the 
$2.5\;\sigma$ confidence level, suggesting that the 3C 295 overdensity 
of sources in the NE chip is truly peculiar.
An intriguing explanation, to be confirmed when the redshifts of
the sources are measured, is that the overdensity of sources is 
actually related to a cosmic filament of the  LSS.

Cappi et al. (2001) discussed four possible causes for this `surplus'
of sources: (1) gravitationally lensed very faint sources; (2) rapid
evolution of cluster AGN or starburst galaxies; (3) cosmic variance of
background sources; (4) LSS associated with the
clusters.  Since the surplus sources are not symmetrically place
around the cluster, our results rule out an enhanced AGN population in
the 3C~295 cluster itself, and lensing by the cluster potential.
Since N-body and hydrodynamical simulations and galaxy surveys
(e.g 2dF and Sloan) lead us to expect that
clusters of galaxies lie at the nexus of several filament, we believe
that this excess is likely to represent a filament of the LSS 
of the Universe converging onto the 3C 295 cluster.  

The redshifts of the sources making up the excess are not yet
known. However, if we assume that the
sources are associated with the 3C 295 cluster, then we can use the
cluster redshift to estimate some of their properties. 

Adopting a concordance cosmology ($H_0 =65$ km/s Mpc, $\Omega_M = 0.3$
and $\Omega_{\Lambda} = 0.7$, Spergel et al. 2003) we obtain
luminosities in the range $7.5\times 10^{41} \div 1.1 \times 10^{44}$
ergs s$^{-1}$ (median $= 3.1 \times 10^{42}$ ergs s$^{-1}$) for the
total {\it Chandra} band.  [$2.8\times 10^{41} \div 3.8\times10^{43}$
ergs s$^{-1}$ (median $= 1.1 \times 10^{42}$ ergs s$^{-1}$) for the
soft band; $1.5\times 10^{42} \div 5.0 \times 10^{43}$ ergs s$^{-1}$
(median $= 4.5 \times 10^{42}$ ergs s$^{-1}$) for the hard band.]
These are moderate, Seyfert galaxy, luminosities, extending down to
the range of luminous starburst galaxies at the faint end of the soft
band.

An association with 3C 295 also defines a spatial scale of $2$ Mpc (5
arcmin) for the adopted cosmology.  For an, admittedly unlikely, spherical region of
volume of $\sim 4 $ Mpc$^{3}$ containing the excess sources, the
implied space density of the excess sources (i.e. subtracting the
CDF-S defined logN-logS contribution) is $0.9$ Mpc$^{-3}$ ($0.5-2$
keV) and $0.8$ Mpc$^{-3}$ ($2-10$ keV). This is well above the normal
maximum AGN space density at $z=0.5$ ($\sim 10^{-4}$ Mpc$^{-3}$, for
luminosities down to $10^{42}$ ergs s$^{-1}$ in the 0.5-2keV band, 
Hasinger 2003, and $10^{43}$ ergs s$^{-1}$ in the 2-10 keV band
Fiore et al. 2003). However,
since the X-ray source luminosities are as low as $\sim 2
\times 10^{41}$ ergs s$^{-1}$ (at the redshift of 3C 295), the
little studied lower end of the XLF is being probed, and contributions
from starburst and even normal galaxies can be important.

In fact, the integral of the field galaxy luminosity function at z=0.5 
(e.g. Poli et al. 2001) gives $\sim 0.13$ galaxies Mpc$^{-3}$
for $M_B < -17$  (a resonable faint end optical luminosity, corresponding to 
our
lower X-ray luminosities). If we
assume that roughly one tenth of the galaxies are active X-ray sources of
$L_X > 3 \times 10^{41}$ ergs s$^{-1}$, then we would expect $\sim0.013$ X-ray
sources Mpc$^{-3}$.
Since we count $\sim0.9$ sources Mpc$^{-3}$, this implies a galaxy
overdensity of $\approx70$, with of course a large (factor of 2-4) positive
and negative uncertainty, because of the uncertainties 
in our space densities and
assumptions. Still, this is intriguingly close to the expected
galaxy overdensity of filaments $\sim 10 \div 10^2$,
and much smaller than the 
overdensities of clusters of galaxies ($\sim 10^3 \div  10^4$), 
(the density contrast at the virial radius being
$\sim200$). 
Furthermore, in a filamentary structure the implied space density will
depend strongly on the orientation of the filament to our line of
sight.

We stress again that these arguments hold only under the assumption that
the redshift of the excess sources is the same of that of the 
3C 295 cluster. In order to determine if this excess is associated 
or not to the 3C 295 cluster, we need to know the redshift of the
sources via optical identifications.

For this reason, in order to study this candidate filament, we are 
pursuing new {\it Chandra} and optical observations 
to map  out the 3C~295 region and delimiting the
filament properties up to scales of $\sim 24$ arcmin (i.e. $\sim 6$
Mpc) from the 3C 295 cluster. These studies are important because they 
may open-up a new way to map high-density peaks of LSS 
at high redshifts with high efficiency.

\begin{acknowledgements}

We thank Dong-Woo Kim for careful reading of the manuscript and for
useful comments. We are in debt with Salvo Sciortino, Giusi Micela
and Francesco Damiani for providing us with the PWdetect algorithm
and the relevant know how.
\end{acknowledgements}




\begin{thebibliography}{}

\bibitem[]{} Almaini, O., Scott, S.E., Dunlop, J.S. et al. 2003, MNRAS, 338, 303

\bibitem[]{}Barrow, J.D., Bhavsar, S.P. \& Sonoda, D.H. 1984, MNRAS, 210, 19

\bibitem[]{}Best, P.N., van Dokkum, P.G., Franx, M \& Rottgering,
H.J.A., 2002, MNRAS, 330, 17

\bibitem[]{}Cappi, M., Mazzotta, P., Elvis, M. et al. 2001, ApJ, 548, 624

\bibitem[]{}Daddi, E., Broadhurst, T., Zamorani, G., Cimatti, A., Roettinger, H.,
\& Renzini, A., 2001, A\&A  376, 825

\bibitem[]{}Damiani, F., Maggio, A., Micela, G. \& Sciortino, S.,
1997a, ApJ, 483, 350

\bibitem[]{}Damiani, F., Maggio, A., Micela, G. \& Sciortino, S.,
1997b, ApJ, 483, 370

\bibitem[]{}Dav\'e, R., Spergel, D.N., Steinhardt, P.J. \& Wandelt,
  B.D., 2001, ApJ, 547, 574

\bibitem[]{}Dickey, J.M. \& Lockman, F.J., 1990, Ann. Rev. Ast. Astr.,ApJ...301..689H
28, 215.

\bibitem[]{}Dressler, A. \& Gunn, J.E., 1992, ApJS, 78, 1

\bibitem[]{}Fasano, G. \& Franceschini, A. 1987, MNRAS, 225, 155

\bibitem[]{}Fiore et al. 2003, A\&A, submitted

\bibitem[]{}Fiore F., Nicastro, F., Savaglio, S., Stella, L. \&  Vietri, M.,
2000, ApJL, 544, 7 

\bibitem[]{}Garmire, G.P., 1997, AAS, 29, 283

\bibitem[]{}Giacconi, R., Zirm, A., Wang, J, et al. 2002, Apj, 139, 369

\bibitem[]{}Giacconi, R., Rosati, P., Tozzi, P. et al. 2001, Apj, 551, 624

\bibitem[]{}Giavalisco, M. \& Dickinson, M., 2001, ApJ, 550, 177 

\bibitem[]{}Gilli, R., Cimatti, A., Daddi, E. 
 et al. 2003, Accepted for pubblication in Apj 
(astro-ph/0304177)

\bibitem[]{}Hasinger, 2003, astro-ph/0302574

\bibitem[]{}Hasinger, G., Burg, R., Giacconi, R., Schmidt, M.,
Trumper, J. \& Zamorani, g., 1998, A\&A, 329, 482


\bibitem[]{} Henry, J.P. \& Henriksen, M.J., 1986, ApJ, 301, 689

\bibitem[]{}Martini, P., Kelson, D.D., Mulchaey, J.S. \& Trager, S.C.,
2002, ApJL, 576 109

\bibitem[]{}Mathur, S., Weinberg, D.H. \& Chen, X.
2003, ApJ, 582, 82

\bibitem[]{}Molnar, S.M., Hughes, J.P, Donahue, M. \& Joy, M., 2002,
ApJL, 573, 91

\bibitem[]{}Nicastro, F., Zezas, A., Drake, J. et al., 2002, ApJ, 573, 157

\bibitem[]{}Nicastro, F., Zezas, A.,  Elvis, M. et al., 2003, Nature,
  421, 719

\bibitem[]{}Peacock, J.A. 1983, MNRAS, 202, 615

\bibitem[]{}Peacock, J.A. 1999, ``Cosmological physics'' (Cambridge:
Cambridge University Press)

\bibitem[]{}Peebles, P.J.E., 1980, `` The Large Scale Structure of the
Universe'' (Princeton: Princeton University Press)

\bibitem[]{}Pentericci, L., Kurk, J.D., Carilli, C.L., Harris, D.E.,
Miley, G.K. \& Rottgering, H.J.A., 2002, astro-ph/0209392

\bibitem[]{}Poli, F.,  Giallongo, E., Fontana, A.,
  Cristiani, S. \& D'Odorico, S., 2001, ApJL, 551, 45

\bibitem[]{}Poli, F., Giallongo, E., Fontana, A., et al. 2003, ApJL, 593, 1

\bibitem[]{}Rosati, P., Tozzi, P., Giacconi, R. et al., 2002, ApJ, 566, 667

\bibitem[]{}Soltan, A.M., Freyberg, M.J. \& Hasinger, G., 2002, A\&A, 395, 475

\bibitem[]{}Tozzi, P., Rosati, P., Nonino, M. et al., 2001, ApJ, 562, 42

\bibitem[]{}Vikhlinin, A. \& Forman, W., 1995, ApJL, 455, 109 

\bibitem[]{}Weisskopf, M.C.,, O'Dell, S.L. \& Van Speybroeck, L.P.,
1996, Proc. SPIE, 2805, 2

\bibitem[]{}Yang, Y., Mushotzky, R.F., Barger, A.J., Cowie, L.L,
  Sanders, D.B., \& Steffen, A.T., 2003, ApJL, 585, 85


\bibitem[]{}Zappacosta, L., Mannucci, F, Maiolino, R. et al., 2002, A\&A, 394, 7

\end{thebibliography}
\end{document}